\documentclass[aps,prd,twocolumn,tightenlines,nofootinbib,showpacs,superscriptaddress]{revtex4-1}
\usepackage[utf8]{inputenc}
\usepackage{amsmath}
\usepackage{amsfonts}
\usepackage{amssymb}
\usepackage{graphicx}
\usepackage{rotating}
\usepackage{epsfig,float}
\usepackage{color}
\usepackage{enumitem} 
\usepackage{hyperref}
\hypersetup{backref,colorlinks=true,urlcolor=blue,linkcolor=blue,citecolor=blue}
\usepackage[dvipsnames]{xcolor}

\newcommand{\eps}{\epsilon}
\newcommand{\non}{\nonumber\\}

\begin{document} 
\title{The pole structure of the $\boldsymbol{\Lambda(1405)}$ in a recent QCD simulation}

\author{R. Molina}
\email{ramope71@email.gwu.edu}
\affiliation{Department of Physics, The George Washington University, 725 21$^{\rm st}$ St NW, Washington, DC 20052, USA}

\author{M. D\"oring}
\email{doring@gwu.edu}
\affiliation{Department of Physics, The George Washington University, 725 21$^{\rm st}$ St NW, Washington, DC 20052, USA}
\affiliation{Thomas Jefferson National Accelerator Facility, 12000 Jefferson Ave, Newport News, VA 23606, USA}

\begin{abstract}
  The $\Lambda(1405)$ baryon is difficult to detect in experiment, absent in
many quark model calculations, and supposedly manifested through a two-pole
structure. Its uncommon properties made it subject to numerous experimental and
theoretical studies in recent years. Lattice-QCD eigenvalues for different quark
masses were recently reported by the Adelaide group. We compare these
eigenvalues to predictions of a model based on Unitary Chiral Perturbation
Theory. The U$\chi$PT calculation predicts the quark mass dependence remarkably
well. It also predicts the overlap pattern with different meson-baryon
components, mainly $\pi\Sigma$ and $\bar KN$, at different quark masses, which might help in the construction of meson-baryon operators for improved level detection on the lattice. More
accurate lattice QCD data are required to draw definite conclusions on the
nature of the $\Lambda(1405)$.
\end{abstract}
\maketitle


\section{Introduction}
The $\Lambda(1405)$ has been a controversial state for many years. In the quark
model it is classified as $q^3$ state belonging to the $70$- dimensional
representation with excitation of one of the quarks to the $p$ state
\cite{Isgur:1978xj}. A pentaquark structure $q^4\bar{q}$ has also been proposed
\cite{Inoue:2006nf}. Nevertheless, the mass of the $\Lambda(1405)$, that is
lighter than the $N(1535)$, and the large spin-orbit splitting between the
$\Lambda(1405)$ and $\Lambda(1520)$ were difficult to understand in the quark
model picture. The $\Lambda(1405)$ has been considered  as a quasibound
molecular state of the $\bar{K}N$ system for many years \cite{Dalitz:1960du},
\cite{Dalitz:1967fp}. In fact, there are experimental evidences that  the
$\Lambda(1405)$ resonance, which has been observed in the $\pi\Sigma$ invariant
mass distribution, is mostly  a $\bar{K}N$ and/or $\pi\Sigma$ composite
\cite{Veit:1984jr,Siegel:1994mb, Tanaka:1992gj,Chao:1973sa, confedalitz}. The
reason is that the $\Lambda(1405)$ lies just $25$ MeV below the $\bar{K}N$
threshold and has a strong influence in the low-energy $\bar{K}N$ data
\cite{Veit:1984jr,Siegel:1994mb,Chao:1973sa,confedalitz}.   It should be
stressed, however, that the partial-wave content, and in particular the
$S$-wave, is difficult to determine close to threshold as demonstrated recently
by the ANL/Osaka~\cite{Kamano:2014zba, Kamano:2015hxa} and Kent State
groups~\cite{Zhang:2013cua}. See also a recent re-analysis of the  KSU partial
waves to extract the resonance content \cite{Fernandez-Ramirez:2015tfa}. Better
kaon-induced reaction data are needed~\cite{Jackson:2015dva, magasan,Briscoe:2015qia}.

The $\Lambda(1405)$ also played an important role in the so-called kaonic
hydrogen puzzle~\cite{Gasser:2007zt, Veit:1984jr, Jennings:1986yg, Arima:1990yv,
Arima:1994bv, He:1993et,Siegel:1994mb,Kumar:1980hs,Schnick:1987is, Fink:1989uk,
Tanaka:1992gj}, which was resolved through accurate measurements of the $1S$
level shift of the kaonic hydrogen atom from atomic X-rays
\cite{Iwasaki:1997wf,Ito:1998yi}. From these measurements, the  $K^-p$
scattering length can be extracted (through the Deser Formula
\cite{Deser:1954vq}). A precise determination of the $K^-p$ scattering length requires to
include isospin breaking corrections~\cite{Meissner:2004jr}.
   
The accurate kaonic hydrogen measurements by DEAR and
SIDDHARTA \cite{Bazzi:2011zj, DEARSID}, together with total cross section data and threshold
branching ratios, are successfully described in the framework of chiral SU(3)
coupled-channels dynamics with input based on the NLO meson-baryon effective
Lagrangian \cite{Borasoy:2005ie, Oller:2005ig, Oller:2006jw,
Ikeda:2012au,Mai:2012dt, Guo:2012vv, Kamiya:2016jqc, Cieply:2016jby}. Dispersion relations can be used to perform the necessary resummation of
the chiral perturbation theory amplitudes at any order \cite{Oller:2000fj}.
In particular, the kaonic hydrogen data are used to constrain the meson-baryon coupled-channel
amplitudes, giving rise to a more precise determination of  the location of the
two poles. Off-shell effects in the NLO chiral expansion of the effective
Lagrangian lead only to small changes of the pole
positions \cite{Mai:2012dt}. Implications of the new data for $\bar{K}d$
scattering are discussed in Refs.~\cite{Doring:2011xc, Mai:2014uma}. See
Ref.~\cite{Cieply:2016jby} for a recent comparison of approaches.

Since the $\Lambda(1405)$ mass lies between the $\pi\Sigma$ and $\bar{K} N$
thresholds, a coupled-channel description is mandatory. In fact, all the unitary
frameworks based on chiral Lagrangians for the study of the $S$- wave
meson-baryon interaction lead to the generation of  this resonance
\cite{Kaiser:1996js,Oller:2000fj, bennhold,Kaiser:1995eg, Oset:1997it,
Jido:2002yz, Jido:2003cb, GarciaRecio:2002td, Ikeda:2012au, Mai:2012dt}.
Within the $U\chi PT$ framework, two poles close to the $\Lambda(1405)$
resonance mass appear \cite{Oller:2000fj,Jido:2003cb}. This was also the case in
the  cloudy bag model of Ref. \cite{Fink:1989uk}. The coupled-channel formalism
takes into account all possible ($J^P=\frac{1}{2}^-;I=0$) pseudoscalar
meson-octet baryon channels  (except for $\eta'\Lambda$ whose coupling is
supposed to be negligible): $\bar{K}N$, $\pi \Sigma$ , $\eta \Lambda$ and
$K\Xi$  \cite{GarciaRecio:2002td, Oset:1997it, Doring:2010rd}. For example, in
Ref.~\cite{Doring:2010rd} the two states are found in the complex plane of
scattering energy at $\sqrt{s}=(1390-66\,i)$~MeV  and $(1426-16\,i)$~MeV. Both
states lie on the same Riemann sheet, with the real parts of their pole
positions above the $\pi\Sigma$ and below the $\bar K N$ threshold. In most
approaches, the lower state is wider and couples stronger to the $\pi\Sigma$
channel, while the upper state close to the $\bar K N$ threshold is narrower and
couples more to $\bar K N$.  The position and width of the lighter state is less
well determined than for the heavier state~\cite{Borasoy:2006sr, Mai:2012dt}.

Evidence of the proposed two-pole structure~\cite{Oller:2000fj, Jido:2003cb} has
been accumulated through the study of different reactions. For instance, in Ref.
\cite{Geng:2007vm}, the theoretical study of the $pp \to p K^+ \Lambda(1405)$
reaction shows different line-shapes from the $K$, $\pi$ and $\rho$-exchange
contributions due to the two-pole structure, and its sum is consistent with
experimental data. Indeed, the two poles associated with the $\Lambda(1405)$ can
also be studied by means of different production reactions which favor one or
the other pole.  In Ref. \cite{Magas:2005vu}, it is shown that the $K^-p \to
\pi^0 \pi^0 \Sigma^0$ reaction is sensitive to the second pole of the
$\Lambda(1405)$ resonance. In this process the $\pi^0$ is emitted prior to the
$K^- p\to \pi ^ 0 \Sigma ^0$ reaction, which gives more weight to the second
state. The model of Ref. \cite{Magas:2005vu} reproduces both the invariant mass 
distributions and integrated cross sections observed in the experiment by
the Crystal Ball Collaboration \cite{Prakhov:2004an}. Other reactions to unravel
the two-pole structure of the $\Lambda(1405)$ have been proposed in Ref.
\cite{Geng:2007hz}. The $\pi^0\Sigma^0$ decay mode of the $\Lambda(1405)$ is, in
general, clean because there is no contamination from the $\Sigma(1385)$. In
contrast to the reaction $K^-p \to \pi^0 \pi^0 \Sigma^0$, the reaction $\pi^-
p\to  K^0 \pi \Sigma $ studied in Ref.~\cite{Hyodo:2003jw}  shows a different
shape of the resonance, and is dominated by the $\pi \Sigma \to \pi \Sigma$
amplitude, hence, favoring the lower and wider state. Further evidence for two
$\Lambda(1405)$ states is found in Refs. \cite{Hyodo:2004vt,Jido:2009jf}. The
composite nature of the $\Lambda(1405)$ as $\bar{K}N$ bound state has been
investigated in Refs. \cite{Sekihara:2014kya,Hyodo:2008xr}. Regge
trajectories of the two poles of the $\Lambda(1405)$ have been studied in Ref.
\cite{Fernandez-Ramirez:2015fbq}.

Recently, the spin and parity of $\Lambda(1405)$ were deduced based on $\gamma 
p \to K^+ \Lambda(1405)$ reaction data~\cite{Moriya:2013eb} measured at CLAS,
and confirmed to be $1/2^-$ \cite{Moriya:2014kpv}. The lineshape of the
$\Lambda(1405)$ differs in the $\pi^{+}\Sigma^{-}$ and $\pi^{-}\Sigma^{+}$ decay
channels as a result of the isospin interference between different $\pi\Sigma$
channels.   In Refs.~\cite{Mai:2014xna, Roca:2013av, Roca:2013cca} the impact of
the new photoproduction data~\cite{Moriya:2013eb} on the pole structure of the
$\Lambda(1405)$ has been quantified. 

The finite-volume spectrum of the $\Lambda(1405)$ was predicted in
Ref.~\cite{Doring:2011ip} based on a dynamical coupled-channel model and a
chiral unitary approach. The coupled-channel $\bar KN$, $\pi\Sigma$ scattering
lengths in the finite volume were discussed in Ref.~\cite{Lage:2009zv}. The
problem of multiple thresholds and resonances in finite-volume baryon
spectroscopy was discussed for the example of the $N(1535), \,N(1650)$ in
Ref.~\cite{Doring:2013glu}. After this manuscript appeared on arXiv, the $\Lambda(1405)$ finite-volume spectrum was analyzed in Ref.~\cite{Liu:2016wxq}.

Recently, the spectrum of excited hyperons became accessible in ab-initio
simulations of QCD on the lattice~\cite{Bulava:2010yg, Menadue:2011pd,
Engel:2012qp, Engel:2013ig, Edwards:2012fx, Melnitchouk:2002eg,
WalkerLoud:2008bp}. The determination of meson-baryon phase shifts has been
pioneered for $S=0,J^P=\frac{1}{2}^-$ in Ref. \cite{Lang:2012db}. 

The aim of the present study is to test the two-pole hypothesis of the
$\Lambda(1405)$ in the light of the new lattice QCD data from Ref.~\cite{hall}.
This work is indeed the first comparison between lattice data and a prediction from U$\chi$PT.
For this, we determine the $M_\pi^2$-evolution of the eigenvalues with $I=0$,
$S=-1$ and $J^P=1/2^{-}$, using the lowest order chiral interaction in the
finite volume, for several sets of ground state masses, in particular, the
physical set, and the sets of pion masses used in Ref.~\cite{hall} which are between 170 MeV and 620 MeV. We will study the
properties of the first two states, pole positions, distances to the $\bar{K}N$
and $\pi\Sigma$ thresholds, and couplings to the meson-baryon components, and
compare to the lattice data. 

 The article is organized as follows. In Section II we describe the
 coupled-channel formalism using the U$\chi$PT lowest order potential, to
 dynamically generate the $\Lambda(1405)$. In Section III,  we explain how to
 calculate the bound states in the box using the coupled-channel formalism.
 Finally, in Section IV and V we present results and conclusions.


\section{The $\mathbf{\Lambda(1405)}$ in the infinite volume}

In the chiral unitary approach, the $\Lambda(1405)$ resonance is  dynamically
generated in $s$-wave meson-baryon scattering from the coupled channels with isospin $I=0$, and strangeness $S=-1$,
$\bar{K}N$, $\pi\Sigma$, $\eta\Lambda$ and $K\Xi$. The scattering equation used
to study the meson-baryon system is \cite{Oller:2000fj}
\begin{equation}
   T = (1-VG^{DR})^{-1}\,V\ ,
   \label{bse}
\end{equation}
where the matrix $V$ is the interaction kernel of the scattering equation, in
$s$-wave given by the lowest order of chiral  perturbation theory (the
Weinberg-Tomozawa interaction),
\begin{eqnarray}
&&V_{i j} (W) =
 - C_{i j} \frac{1}{4 f_i f_j}(2W - M_{i}-M_{j}) \non
&&\times
\sqrt{\frac{M_{i}+E_{i}}{2M_{i}}}
\sqrt{\frac{M_{j}+E_{j}}{2M_{j}}}
\label{eq:ampl2}
\end{eqnarray}
with the channel indices $i,j$, the baryon masses $M$,  the meson decay
constants $f$, the baryon on-shell energy $E$ and the center of mass energy $W$
in the meson-baryon system.  The coefficients $C_{ij}$ are the couplings
strengths to the pseudoscalars ($P$) and baryons ($B$) of each reaction
$P_iB_i\to P_jB_j$ ($i,j=1,\dots, 4$), determined by the lowest-order chiral
Lagrangian in isospin $I=0$, 
\begin{equation}
C=\begin{pmatrix}
3			& -\sqrt{\frac{3}{2}}	& \frac{3}{\sqrt{2}}	& 0			\\
-\sqrt{\frac{3}{2}}	& 4			& 0			& \sqrt{\frac{3}{2}}	\\
\frac{3}{\sqrt{2}}	& 0			& 0			& -\frac{3}{\sqrt{2}}	\\
0			& \sqrt{\frac{3}{2}}	& -\frac{3}{\sqrt{2}}	& 3
\end{pmatrix} \ .
\end{equation} 
All calculations are performed in the isospin limit. The precision analysis of
experimental data requires to take isospin breaking into account, especially for
kaonic hydrogen \cite{Meissner:2004jr}. However, the lattice data of Ref.
\cite{hall} to which we will compare neglect isospin breaking as well. 

The diagonal matrix $G^{DR}_i$ in Eq. (\ref{bse}) is the meson baryon loop
function, evaluated using dimensional regularization as
\cite{Oller:2000fj}
\begin{eqnarray}
&&G_i^{DR}(W)
= i \,  \int \frac{d^4 q}{(2 \pi)^4} \,
\frac{2 M_i}{q^2 - M_i^2 + i \eps} \, \frac{1}{(P-q)^2 - m_i^2 + i
\eps}\non
 &&= \frac{2 M_i}{16 \pi^2} \left\{ a_i(\mu) + \ln
\frac{M_i^2}{\mu^2} + \frac{m_i^2-M_i^2 + W^{2}}{2W^{2}} \ln \frac{m_i^2}{M_i^2} 
\right.  \nonumber\\&&+
\frac{ q_{\mathrm{cm}}}{W}
\left[
\ln(\hspace*{0.2cm}W^{2}-(M_i^2-m_i^2)+2  q_\mathrm{cm} W)
\right. \nonumber\\
 &&+\ln(\hspace*{0.2cm}W^{2}+(M_i^2-m_i^2)+2 q_\mathrm{cm} W)\nonumber\\&&-  \ln(-W^{2}+(M_i^2-m_i^2)+2 q_\mathrm{cm} W)  \nonumber\\&&\left.\left.- \ln(-W^{2}-(M_i^2-m_i^2)+2 q_\mathrm{cm} W) \right]
\right\} ,
\label{eq:gpropdr}
\end{eqnarray}
where $m$ are the meson masses,
$q_{\mathrm{cm}}$
is the three-momentum of the meson or baryon  in the center-of-mass frame and  $\mu$ is the
scale of dimensional regularization chosen as $\mu=630$ MeV in Ref.~\cite{Oset:1997it}. The remaining
finite constants denoted by $a_i(\mu)$ are determined  phenomenologically by a
fit in order to reproduce the threshold branching ratios of $K^{-}p$ to
$\pi\Lambda$ and $\pi\Sigma$ observed   by stopped $K^{-}$ mesons in
hydrogen~\cite{Tovee:1971ga,Nowak:1978au}.
The $a_{i}$ constants were determined in Ref.~\cite{bennhold} using the same
averaged decay constant for all the pseudoscalar mesons involved,
$f=1.123\,f_\pi$. The latter relation changes for unphysical pion masses. Thus, it is more appropiate to use different decay constants $f_i,f_j$ in
Eq. (\ref{eq:ampl2}) depending on which mesons are  in the external legs of the
pseudoscalar-baryon interaction, $P_i B_i\to P_j B_j$. The decay constants $f_\pi,\,f_K,\,g_\eta$ are obtained for unphysical masses using the SU(3) chiral unitary
extrapolation of Ref.~\cite{jenifer} as discussed in the Appendix. That extrapolation was obtained in a fit to decay constants on the lattice at different pion masses. The
subtraction constants found here are $a_{\bar K N} = -2.2$, $a_{\pi \Sigma} = -
1.6$, $a_{\eta \Lambda} = -2.5$, $a_{K \Xi} = -2.9$. These values are chosen to produce almost identical
amplitudes as in Ref.~\cite{bennhold} for physical pion masses.  In addition, these values are close to a
natural value equivalent to the three-momentum cut-off of $630$
MeV~\cite{Oller:2000fj}. The former produce the same description of scattering cross sections and threshold branching ratios as in the original Ramos/Oset paper~\cite{Oset:1997it}. The reader is referred to that study for pictures of cross sections and their description by the model.

It should be stressed that the present model allows for an exploratory and qualitative study of lattice QCD eigenvalues. The lattice data discussed later are sparse and have large uncertainties compared to the experimental uncertainties. Yet, as discussed in the Introduction, new experimental data have been produced that are contained in the most recent analyses~\cite{Mai:2012dt, Guo:2012vv, Ikeda:2012au, Kamiya:2016jqc, Cieply:2016jby}. An update of the present results, using one of these more quantitative studies would allow to study the impact of experimental data on the finite-volume predictions performed here, and also to improve the chiral extrapolation as most of the newer models contain next-to-leading order contributions. For this to provide new insights, the precision of the lattice data should also improve.

The amplitudes $T_{ij}$ can be analytically continued along the right-hand cut
into the lower $W$ plane (Im $W<0$) by substituting (index $DR$ omitted)
\begin{eqnarray}
 G^{II}_i(W)=\begin{cases}
 G_i(W)+i\,\frac{2M_i\, q_{\mathrm{cm}}}{4\pi W}\ ,&  
 {\rm for\,\, Re}\,W>m_i+M_i\nonumber\\
 G_i(W)\ , & {\rm else} \end{cases}
 \label{eq:gsec}
 \end{eqnarray}
in Eq. (\ref{bse}) to ensure that the resonance poles closest to the physical
axis are searched for. The residua $a_{-1}^{ij}$ of the poles factorize
channel-wise, $a_{-1}^{ij}=g_ig_j$, defining the coupling strengths $g_{i}$  of
the resonance to the meson-baryon channels.  The scattering amplitude  for the
channels $i$ and $j$ close to the resonance pole at $W=W_0$ can be approximated
as $T_{ij}\simeq g_i g_j/(W-W_0)$. As in
Refs.~\cite{Oller:2000fj,Jido:2002yz,Jido:2003cb} the amplitude in the present
study exhibits two poles at $W_0 = (1379 -71i)$ and $(1412 -20i)$ MeV. Both
poles are situated on the same Riemann sheet. As the size of the couplings in
Table~\ref{tab:coupl} shows, the lighter state couples predominantly to the
$\pi\Sigma$ channel while the heavier state couples stronger to the $\bar KN$
channel. If the transitions between these channels are set to zero, the lighter
state is still present as a resonance in the $\pi\Sigma$ channel while the
heavier state becomes a bound state in the $\bar KN$ channel. This demonstrates
that each pole can be understood as dynamically generated from the respective
channel. The pole position of the $\Lambda(1670)$ is obtained here at $W_0 =(
1672 -18i )$ MeV. It appears as a quasi-bound $K\Xi$ state as the large coupling
in Table~\ref{tab:coupl} indicates.
\begin{table*}
\hspace{-0.25cm}\scalebox{0.91}{{\setlength{\tabcolsep}{0.15em}
{\renewcommand{\arraystretch}{1.5}
\begin{tabular}{ccccccc}
 \hline
\hline
  & $\bar{K}N$ & $\pi\Sigma$ & $\eta\Lambda$ & $K\Xi$ &$1-\cal{Z}$&$\cal{Z}$ \\\hline
\multicolumn{7}{c}{$W_0=1379 -71i$}\\
 $g_i$  $(|g_l|)$& $-0.9+2.0i$  $(2.2)$&$2.4-1.9i$  $(3.1)$&$0.06+0.8i$  $(0.8)$& $0.3-0.4i$  $(0.5)$&&\\
${\cal{P}}_l$  $(|{\cal{P}}_l|)$&$-0.23-0.05i$  $(0.23)$&$0.52+0.53i$  $(0.74)$&$-0.014+0.005i$  $(0.014)$&$-0.002-0.004i$  $(0.005)$&$0.28+0.47i$&$0.72-0.47i$\\
\multicolumn{7}{c}{$W_0=1412 -20i$}\\
  $g_l$  $(|g_l|)$&$3.0+0.7i$  $( 3.1)$&$-0.9-1.5i$  $(1.7)$&$1.5+0.08i$  $(1.5)$&$-0.2-0.3i$  $(0.3)$&&\\
${\cal{P}}_l$  $(|{\cal{P}}_l|)$&$0.92-0.0098i$  $(0.92)$&$-0.15-0.15i$  $(0.21)$&$0.05+0.002i$  $(0.05)$&$-0.0005+0.002i$ $(0.002)$&$0.82-0.16i$&$0.18+0.16i$\\
\multicolumn{7}{c}{$W_0=1672 -18i $}\\
 $g_l$  $(|g_l|)$&$0.4-0.7i$  $(0.8)$ &$0.03+0.3i$  $(0.3)$&$-1.1+0.05i$  $(1.1)$ &$3.3-0.16i$  $(3.4)$&&\\  
${\cal{P}}_l$ $(|{\cal{P}}_l|)$&{\small{$0.026+0.0037i$  $(0.026)$}}&{\small{$0.0012-0.0028i$  $(0.0031)$}}&$-0.12+0.16i$  $(0.20)$&$0.46-0.089i$  $(0.47)$&$0.37+0.073i$&$0.63-0.073i$\\
  \hline\hline
\end{tabular}}}}
\caption{Coupling constants {$|g_i|$} to the meson-baryon channels obtained as the residua of the scattering amplitude at the pole position $W_0$, and the quantities $\mathcal{P}_l$, $ 1-\mathcal{Z}$ and $\mathcal{Z}$, discussed following Eq. (\ref{eq:compw}).}
\label{tab:coupl}
\end{table*}

\subsection{Compositeness and Elementariness}
 The magnitudes of these couplings provides an idea of the strength of the
coupling between the bound state and the meson-baryon channel. However, it is
known that a coupling $g_i$ to a channel $i$, that opens far above the state,
might be large although that channel is irrelevant for the wave function of the
state. It is therefore more realistic to consider the relative weight $P_l $ of
a channel in the wave function of a state, $\mathcal{P}_l=-\left(g_l^2
\frac{\partial{G}^{II}_l}{\partial W}\right)_{W=W_0}$, which fulfills the
identity  \cite{Weinberg:1962hj, Weinberg:1965zz, Baru:2003qq, Hanhart:2011jz, Sekihara:2010uz, Hyodo:2011qc, Hyodo:2013nka, danioset, acetioset, Aceti:2012dd, ollerguo}, 
 \begin{equation}
  1=-\left(\sum_{l}g_l^2 \frac{\partial{G}^{II}_l}{\partial W} +\sum_{k,l}g_k G_{k}^{II}\frac{\partial{V}_{kl}}{\partial{W}}G^{II}_lg_l\right)_{W= W_0}\ .
 \label{eq:compw}\end{equation}
The above equation can be regarded as the generalized version of the Weinberg compositeness
condition for the coupled-channel case. Usually, the first term on the right
hand side is identified with {\it compositeness} $\mathcal{X}\equiv
1-\mathcal{Z}=\sum_l\mathcal{P}_l$ and $\mathcal{Z}=-\left(\sum_{k,l}g_k
G_{k}^{II}\frac{\partial{V}_{kl}}{\partial{W}}G^{II}_lg_l\right)_{W= W_0}$ is
referred to as {\it elementariness}. These quantities are complex numbers in
general. In the special case of bound states, $\mathcal{X}$ and $\mathcal{Z}$
take real values. For bound states, $0\le\mathcal{X}\le 1$, can be interpreted as the
probability of the state to be in any of the considered channels and $\mathcal{P}_l$ gives the probability of a particular channel $l$ to be
in the wave function of the state
\cite{Sekihara:2014kya,Sekihara:2010uz,danioset,acetioset}. In contrast,
$\mathcal{Z}$, which can be directly related to the derivative of the potential
with respect to the energy, gives the probability that the state overlaps with 
a channel not explicitly contained in the amplitude
\cite{Sekihara:2014kya,Sekihara:2010uz,danioset,acetioset}. When these
quantities take complex values (as for resonances), it is not possible to
interpret them as probabilities but these magnitudes are rather extrapolations
of probability in the complex plane  of the energy \cite{acetioset}. The
first term on the right hand side of Eq. (\ref{eq:compw}), 
$\left(-\sum_{l}g_l^2 \frac{\partial{G}^{II}_l}{\partial W}\right)_{W=W_0}$,
equals $\int d^3 p\langle \vec{p}\vert \Psi \rangle^2$, not $\int d^3
p\vert\langle \vec{p}\vert \Psi \rangle\vert^2$ \cite{acetioset}.  Therefore,
one can still interpret $\mathcal{P}_l$ as a magnitude that provides the
relevance of  a given channel in the wave function of the state.  

The quantities $\mathcal{P}_l$, $1-\cal{Z}$ and $\mathcal{Z}$ are given in Table
\ref{tab:coupl} for the three states obtained in the four-coupled-channel
calculation discussed here. The $\pi\Sigma$ and
$\bar{K}N$ channels are relevant in the case of the two poles associated to the
$\Lambda(1405)$, while the $\eta \Lambda$ and $K\Xi$ channels have more strength
in the square of the wave function related to the pole of the $\Lambda(1670)$. These
results are in line with previous calculations
\cite{Sekihara:2010uz,acetioset,Garcia-Recio:2015jsa}. 

However, how to interpret the \textit{elementariness} and \textit{compositeness}
for resonances is still controversial. Because the imaginary parts cancel in Eq.
(\ref{eq:compw}), the authors of Refs. \cite{acetioset,Garcia-Recio:2015jsa},
reinterpret $1-Z\equiv\mathrm{Re}(1-\cal{Z})$ ($=\mathrm{Re}\int d^3 p\langle
\vec{p}\vert \Psi \rangle^2$ \cite{acetioset}), as the compositeness of
resonances (and the same for $Z\equiv\mathrm{Re}\cal{Z}$, which is called the
elementariness). In this interpretation, the lower pole of the $\Lambda(1405)$
has a high elementariness, while the second pole is interpreted as mainly
$\bar{K}N$ composite. Nevertheless, in Ref. \cite{ollerguo} a new interpretation
of these magnitudes is proposed, i. e., the compositeness is reinterpreted as
$X\equiv\sum_l|g_l|^2\left|\frac{\partial G_l(W)}{\partial W} \right|_{W=W_0}$.
Within the criterion of Ref. \cite{ollerguo} both poles of the $\Lambda(1405)$
would be $\pi \Sigma-\bar{K}N$ composites. Other attempts to define the
concepts of \textit{elementariness} and \textit{compositeness} in terms of real
quantities that can be associated to probabilities have been done in Ref.
\cite{Sekihara:2015gvw}. In any case, the two poles of the $\Lambda(1405)$ and
also the $\Lambda(1670)$ emerge from the unitarization of the lowest-order,
longest-range interaction. In that sense, these states can be interpreted as
loosely quasi-bound meson-baryon molecules.


\section{Formalism in finite volume }
The loop function $G$ in Eqs. (\ref{bse}), (\ref{eq:gpropdr}) can also be
evaluated with a cutoff \cite{Oller:1998hw}. For channel $i$,
\begin{equation}
G_i(W)=\frac{2 M_i}{(2\pi)^3}\int_0^{q_\mathrm{max}}d^3 q\, I_i(W,\vec{q})\ ,
\label{eq:gfree}
\end{equation} 
with
\begin{equation}
I_i(W, \vec{q})= \frac{\omega^{(i)}_1(\vec{q}) + \omega^{(i)}_2(\vec{q})}{2 \omega^{(i)}_1(
\vec{q})  \omega^{(i)}_2(\vec{q})}\frac{1}{W^2-(\omega^{(i)}_1(\vec{q}) 
+ \omega^{(i)}_2(\vec{q}))^2}\ ,
\label{eq:iq}
\end{equation}
where $\omega^{(i)}_{1} = \sqrt{m^{(i)\,2}+|\vec{q}~|^2}$ and $\omega^{(i)}_{2} = \sqrt{M^{(i)\,2}+|\vec{q}~|^2}$ are the meson and baryon energies.  A
formalism of the U$\chi$PT description in finite volume was introduced in
Ref.~\cite{Doring:2011vk}. Here, we follow the same procedure replacing the
infinite-volume amplitude $T$ by the amplitude $\tilde{T}$ in a cubic box of
size $L$. The finite-volume equivalent of Eq. (\ref{eq:gfree}) reads
\begin{equation}
\tilde{G}_i(W)=\frac{2 M_i}{L^3} \sum_{\vec{q_l}} I_i(W, \vec{q})\ ,
\label{eq:gbox}
\end{equation}
  which is quantized according to
\begin{equation}
\vec{q} = \frac{2\pi}{L} \vec{n}\ ,
\end{equation}
corresponding to the periodic boundary conditions. Here, $\vec{n}$ denotes the
three-dimensional vector of all integers ($\vec n\in\mathbb{Z}^3$).  This form produces a
degeneracy for the set of three integers which have the same modulus, $q^2 =
\frac{4\pi^2}{L^2}m$ (here $q\equiv |\vec{q}|$ and $m$ stands for the natural
numbers). The degeneracy can be exploited to reduce Eq. (\ref{eq:gbox}) to a
one-dimensional summation using the theta-series of a cubic lattice at rest
\cite{Doring:2011ip}. The sum over the momenta is limited by $q_{max}$, such
that $m_{max}=\frac{q_{max} L}{2 \pi}$. As in the infinite volume, the formalism
should be made independent of $q_{max}$ and be related to $a(\mu)$, the
parameter of the dimensional regularization function loop, $G^{DR}$. This is
done in Ref.~\cite{MartinezTorres:2011pr}, obtaining
\begin{eqnarray}
&&\tilde{G}_i=G_i^{DR}+2M_i\times\nonumber\\
&&\lim_{q_{max} \rightarrow \infty} 
\left( \frac{1}{L^3}\sum_{q<q_{max}} I_i(W,\vec{q}) 
- \int_{q<q_{max}} 
\frac{d^3 q}{(2\pi^3)} I_i(W,\vec{q}) \right)\nonumber\\&& 
\equiv G_i^{DR} + \lim_{q_{max} \rightarrow \infty} \delta G_i\ ,
\label{eq:gtil}
\end{eqnarray}
where the quantity between parenthesis, $\delta G$, is finite as $q_{max}
\rightarrow \infty$.
The Bethe-Salpeter equation in the finite volume can be written as,
\begin{equation}
\tilde{T}=(I-V \tilde{G})^{-1}V \ \label{eq:tfin}
\end{equation}
and the energy levels in the box in the presence of the interaction $V$ correspond
to the condition
\begin{equation}
\det(1-V\tilde{G})=0.
\label{eq:tpole}
\end{equation}
In a single channel, Eq. (\ref{eq:tpole}) leads to poles in the $\tilde{T}$
amplitude when  $V^{-1}=\tilde{G}$.  As a consequence, an infinite number of
poles is predicted for a particular box size. For one channel, the amplitude $T$
in the infinite volume for the energy levels $(W_j)$ can be written as 
\begin{equation}
T=(\tilde{G}(W_j)-G(W_j))^{-1}.
\label{eq:bs2}
\end{equation}
which is equivalent to the L\"uscher formalism up to exponentially suppressed
corrections \cite{Doring:2011vk}.

In the future, lattice simulations will use meson-baryon operators to extract
the eigenvalues in the $\pi\Sigma$, $\bar{K}N$ system, and a maximal overlap of
these operators with the wave function of the state is needed. It is desirable
to develop a criterion specifying the relevance of a given channel for a
finite-volume eigenvalue.

In the finite volume, the couplings $\tilde{g}_i$ can be formally computed from
the real-valued residua of the amplitude in the pole position (since
$\tilde{T}_{kl}\simeq \tilde{g}_k \tilde{g}_l/(W-W_0)$), close to a pole). Also,
an identity similar to the generalization of the Weinberg compositeness
condition for coupled channels discussed in the previous section, Eq.
(\ref{eq:compw}), can be easily obtained by just replacing the meson-baryon
function loop, $G$, and scattering amplitude, $T$, by their respective functions
in the finite volume, $\tilde{G}$ and $\tilde{T}$, which are given by Eqs.
(\ref{eq:gtil}) and (\ref{eq:tfin}), in Eq. (\ref{eq:compw}), 
\begin{equation}
  1=-\left(\sum_{l}\tilde{g}_l^2 \frac{\partial{\tilde{G}}_l}{\partial W} 
  +\sum_{k,l}\tilde{g}_k \tilde{G}_{k}
  \frac{\partial{V}_{kl}}{\partial{W}}\tilde{G}_l\tilde{g}_l\right)_{W= W_0}\ .
 \label{eq:compwfin}\end{equation}
In the next section we evaluate
$\tilde{P}_l=-\left(\mathrm{\tilde{g}_l^2}\frac{\partial \tilde{G}}{\partial
W}\right)_{W=W_0}$, and
$\tilde{Z}=-\left(\sum_{k,l}\tilde{g}_k\tilde{G}_k\frac{\partial{V}_{kl}}{\partial
W}\tilde{G}_l \tilde{g}_l\right)_{W=W_0}$.  In the infinite volume, the
$P_l$ specify the relative weight of finding the channel $l$ in the
wave function \cite{danioset,acetioset}. Here, we make a conjecture, i.e., that
the $\tilde{P}_l$ carry this meaning over to the poles of the finite-volume
amplitude $\tilde T$ of Eq. (\ref{eq:bs2}), that specify the finite-volume
eigenvalues. Indeed, Eq. (\ref{eq:compw}) has the same form for the poles of
$\tilde T$, In particular, we interpret the quantity $\tilde
P_l=-\left(\mathrm{\tilde{g}_l^2}\frac{\partial \tilde{G}}{\partial
W}\right)_{W=W_0}$ as relevance of channel $l$ for a given finite-volume
eigenvalue. This information could be used in future lattice simulations to
select suitable meson-baryon operators or to extract the lattice eigenvalues. Operators of meson-baryon type
are not used in Ref. \cite{hall}. In the next section, such operators are
discussed.


\section{Results} 
\label{sec:results}
\subsection{Spectrum}
\label{sec:spectrum}
The energy levels in a box are evaluated by means of Eq. (\ref{eq:tpole}). Meson
and baryon masses are taken from the lattice simulation of Ref. \cite{hall}
while the quark mass dependence of $f_\pi,f_K$ and $f_\eta$ (not provided in
Ref. \cite{hall}) are evaluated through the $SU(3)$  chiral extrapolation of
\cite{jenifer} discussed in the Appendix. The resulting decay constants
are shown in Table \ref{tab:par}.

Results are shown in Fig. \ref{fig:compajo} for the first five energy levels
predicted from U$\chi$PT (solid lines). The lattice data of \cite{hall} are shown
as black dots.   They correspond to a size of the box of around $L\simeq 3$ fm
(see Table \ref{tab:par}). For the physical point in this figure we also take
$L=3$ fm. There is good agreement between the U$\chi$PT prediction of the second
energy level and the lattice data for masses below $400$ MeV. For larger masses
there are discrepancies which are discussed later on this section. However, for the two lowest lattice pion
masses, U$\chi$PT predicts an additional level below the $\pi\Sigma$ threshold,
associated with an attractive $\pi\Sigma$ scattering length. The level is not
only present in common U$\chi$PT calculations that all predict an attractive
$\pi\Sigma$ scattering length, it is also found in the finite-volume
version of the dynamical coupled-channel model of Ref. \cite{Doring:2011ip}. In
Ref. \cite{Doring:2011ip}, that represents the first finite-volume
implementation of dynamical coupled-channel models, the attractive $\pi\Sigma$
interaction arises from explicit $t$- and $u$- channel diagrams. This lowest
level is absent in the lattice simulation of Ref. \cite{hall} as Fig.
\ref{fig:compajo} shows. If that finding is confirmed, it represents a serious
challenge for all discussed hadronic models. However, as the lowest state is a
scattering state, maybe it has simply not been detected in Ref. \cite{hall},
which relies on quark operators to extract the finite-volume spectrum. Level
extraction using meson-baryon operators instead of quark operators could help
detecting this scattering state.
Meson-baryon channels that have large overlap with the various eigenstates are
identified later in this section.

In the chiral extrapolation we include the quark mass dependences of the decay constants
$f_\pi,f_K,f_\eta$ but cannot specify the quark mass dependence of the subtraction constants $\alpha_i$. To estimate the
uncertainties from this source we vary each subtraction constant $\alpha$ gradually for increasing
pion masses by  $5\%$, $10\%$, $15\%$, $20\%$ and $25\%$ corresponding to sets 1
to 5 in Table \ref{tab:par} respectively. Also, to account for uncertainties in the chiral extrapolation of the decay constants
$f_\pi$, $f_K$ and $f_\eta$, we vary them by $5\%$, for all sets (since we considered here pion
mass dependence). Fig. \ref{fig:compajo} shows that even with these rather large changes the predicted levels are still less uncertain that the values from the lattice simulation. However, the discrepancy for pion masses larger than $400$ MeV persists. We could attribute this to different sources like the missing NLO in our model, or to the fact that the chiral extrapolation breaks down at high pion masses due to a genuine component, or that the
discrepancies come from other sources intrinsic to the lattice
computation like underestimated errors.


\subsection{Channel dynamics of levels and poles}
\begin{figure}
 \hspace{-1cm}\includegraphics[width=1.0\linewidth]{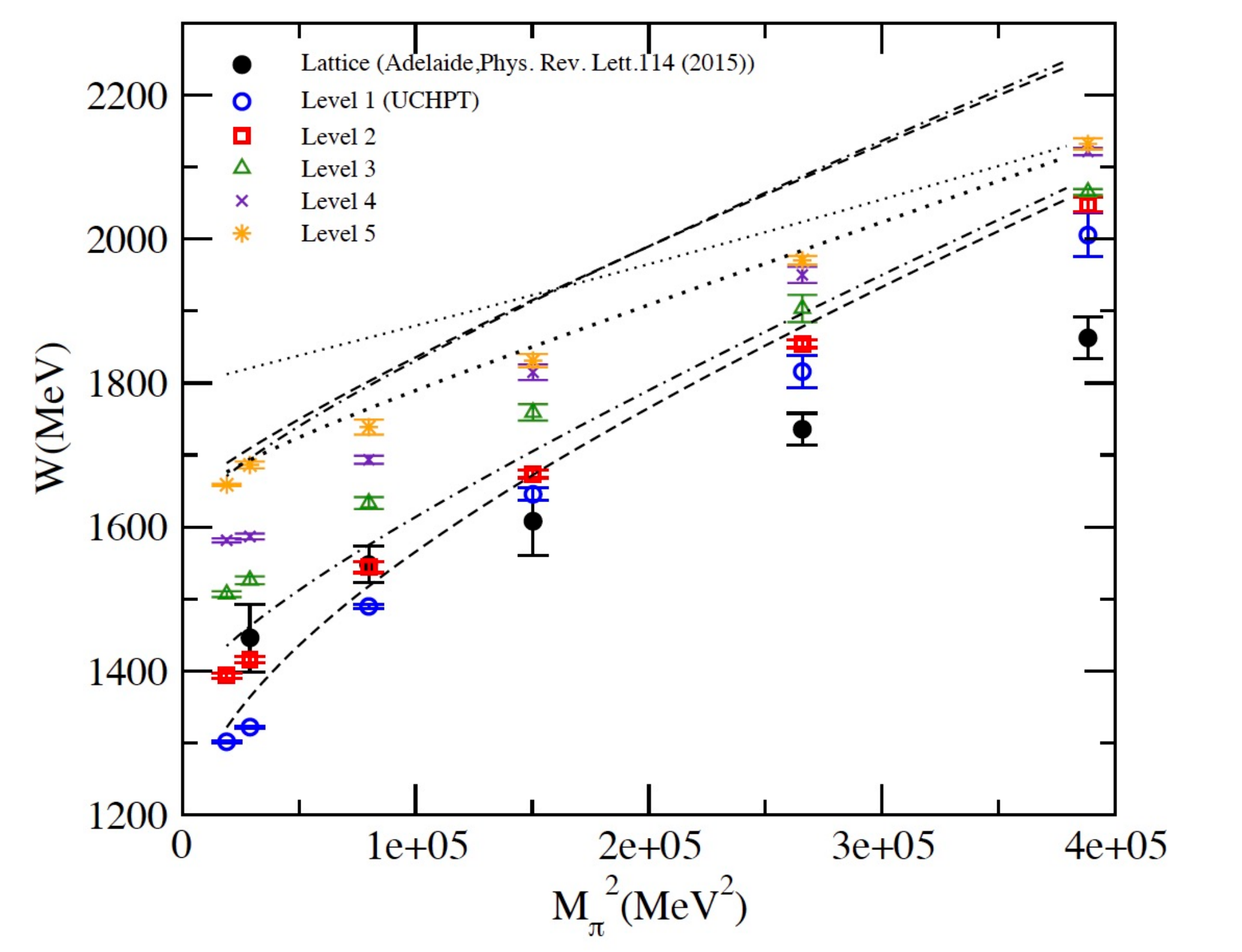}
 \caption{Comparison between the U$\chi$PT prediction for the first five energy
levels (solid lines) and the lattice data of Ref. \cite{hall}, for the physical
set ($L=3$ fm) and sets 1 to 5, as shown in Table \ref{tab:par}. The errors
of our results are obtained by varying subtraction constants and meson decay
constants as described in Sec.~\ref{sec:results}. Here, the dashed, dot-dashed
lines represent $\pi\Sigma$ and $\bar{K}N$ non-interacting levels respectively
(the first two levels are depicted), while the thick-dotted and dotted lines,
show the first $\eta\Lambda$ and $K\Xi$ non-interacting levels, respectively.}
 \label{fig:compajo}
\end{figure}

\begin{figure*}
\includegraphics[width=0.46\linewidth]{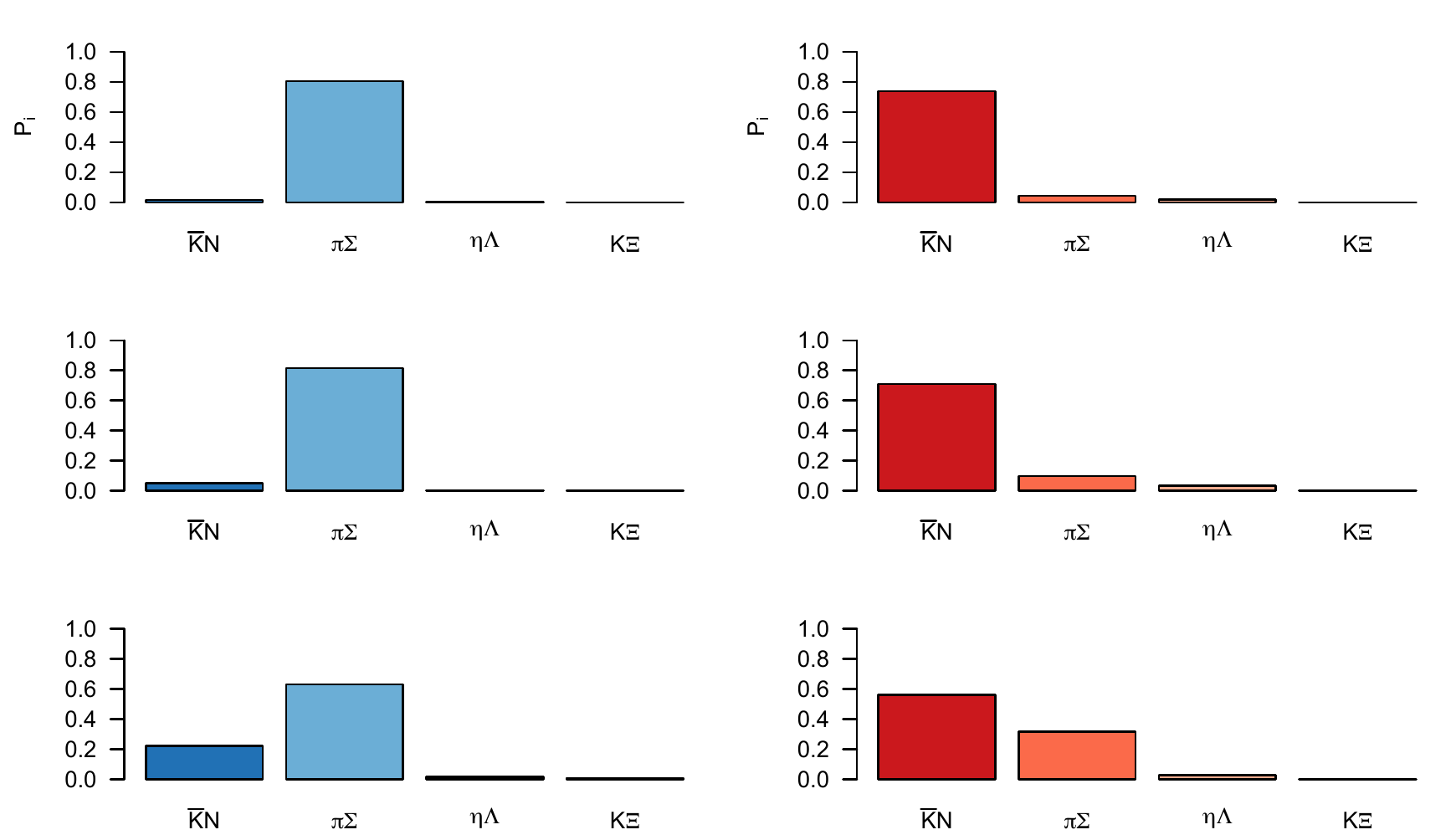}
\includegraphics[width=0.53\linewidth]{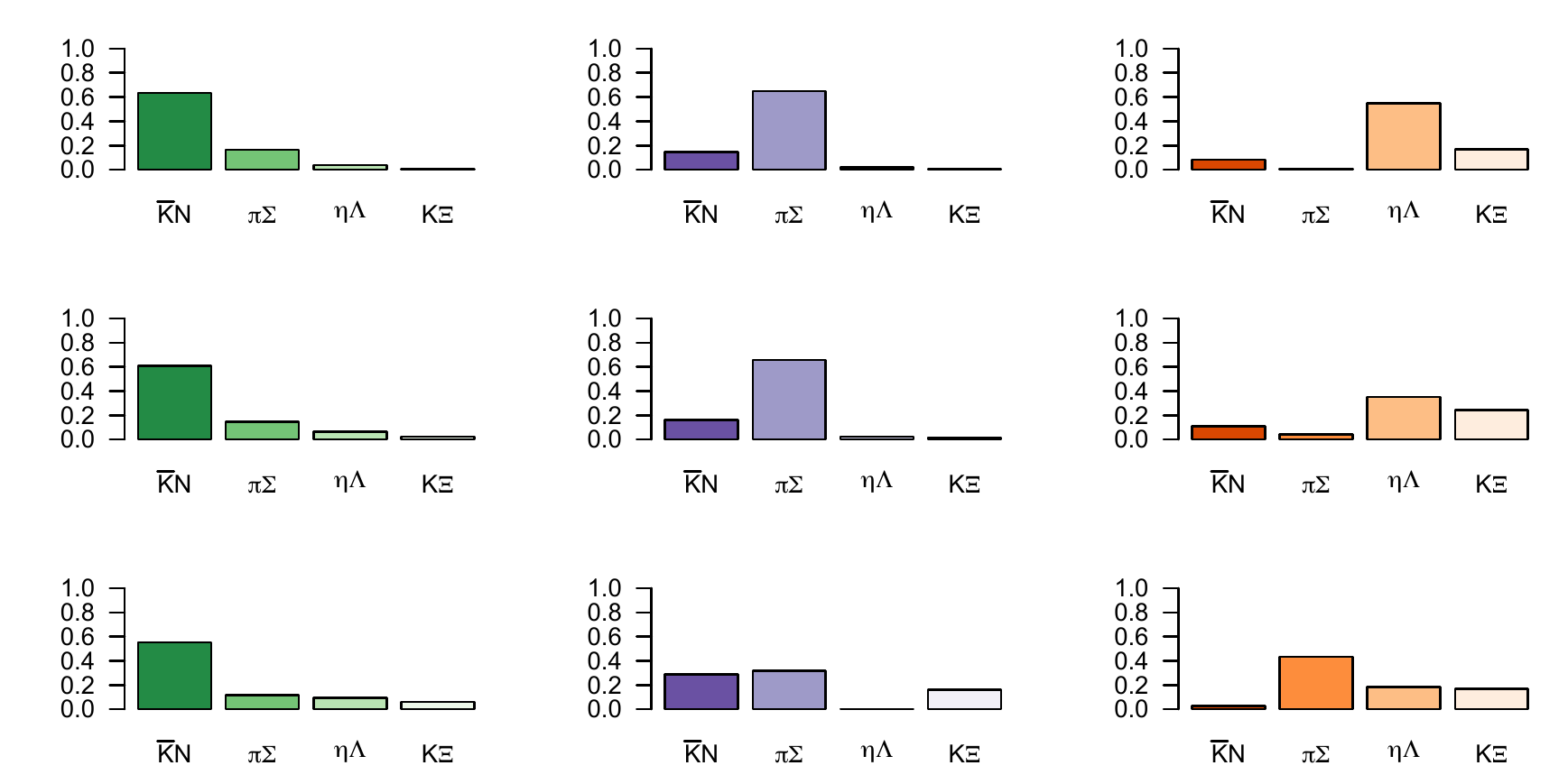}
\caption{Weights of the different channels, $\tilde{P}_i$'s (first term on the
right hand side of Eq. (\ref{eq:compwfin}) for the first five energy levels
(from left to right) and pion masses from $170$ to $388$ MeV (sets 1 to 3 in
Table \ref{tab:par}, from top to bottom).}
\label{fig:lambdacou}
\end{figure*}

In order to understand the role of the meson-baryon channels in the extracted energy
levels, we evaluate the couplings $\tilde{g}_i$, and the magnitudes
$\tilde{P}_i=-\left(\tilde{g}_i^2\frac{\partial \tilde{G}_i}{\partial
W}\right)_{W=W_0}$, $1-\tilde{Z}$, and $\tilde{Z}$, of Eq. (\ref{eq:compwfin})
at the pole position. These quantities are shown in Table \ref{tab:ps} for the
physical mass ($L=3$ fm) and sets 1 to 3 of quark masses shown in Table
\ref{tab:par} (at higher masses the chiral prediction becomes very uncertain and no values are quoted). The part which is related to the energy dependence of the
potential is generally small, $\tilde{Z}\simeq 0-0.3$, and the weights of the
channels $\tilde{P}_i$'s are between $0$ and $1$, like in the infinite volume
for bound states. The $\tilde{P}_i$'s are diagrammatically represented in Fig.
\ref{fig:lambdacou}. Here, the left column of bar diagrams in blue represents
the weights $\tilde{P}_i$ of the lowest energy level, while the following
columns represent the levels $2$ to $5$ with the same color coding as in Fig.
\ref{fig:compajo}. Every level is depicted for pion masses in the range
$170-388$ MeV from top to bottom corresponding to sets $1$ to $3$ in Table
\ref{tab:par}. The $\pi\Sigma$ channel dominates the lowest level. The relative
weights $\tilde{P}_l$ for the $\eta \Lambda$ and $K\Xi$ are almost zero for the
lowest state and for low pion masses (set 1). This confirms the discussed property of
the $\tilde{P}_l$ suppressing effectively the irrelevant channels that open at
much higher energies (compare with the corresponding values for the
$\tilde{g}_i$ in Table \ref{tab:ps}). Also, it is quite natural that the lowest
state has a dominant $\pi \Sigma$ content, as it is a threshold level below the
$\pi\Sigma$ threshold associated with an attractive $\pi\Sigma$ interaction. For
larger quark masses, this trend is inverted and the $\bar{K} N$ strength becomes
larger.  On the other hand, the second energy level (second column in Fig.
\ref{fig:lambdacou}) shows a significant dominance of the $\bar{K}N$ component,
with $\tilde{g}_{\bar{K}N}$ and $\tilde{P}_{\bar{K}N}$ both larger, if the pion
mass is not very high. Although we cannot identify finite-volume energy
eigenstates with resonances, the $\bar KN$ dominance of the second eigenstate is
in line with the second $\Lambda(1405)$ pole being predominantly generated from
the $\bar K N$ channel (cf. Table \ref{tab:coupl}).

From Fig. \ref{fig:compajo}, is clear that the first two lattice data points in
Fig. \ref{fig:compajo} correspond to the second energy level, for which the
$\bar{K}N$ component clearly dominates, while the third lattice data point
could belong to  either the first or second energy level predicted from
U$\chi$PT. For the third energy level, both
$\tilde{g}_i$ and $\tilde{P}_i$ are larger for the $\bar{K}N$ component, while
the $\pi\Sigma$ channel dominates the fourth energy level. 

\begin{figure*}[ht]
 \includegraphics[width=0.7\linewidth]{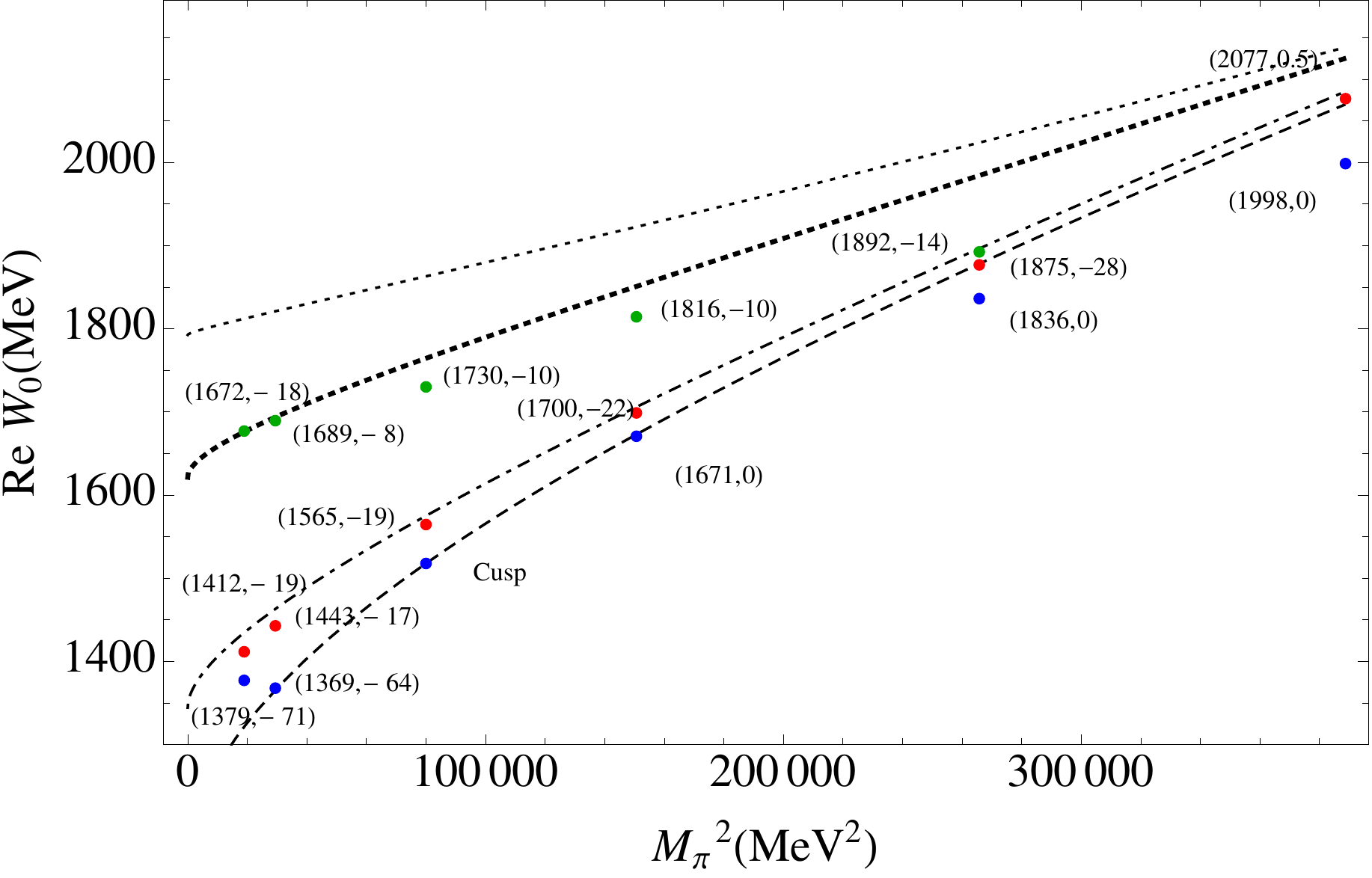}
 \caption{Behavior of the real part of the pole positions, $\mathrm{Re}\,W_0$,
found in the $T$-matrix in the infinite volume with the $M^2_\pi$, for the
physical, and 1 to 5 sets. The numbers in parenthesis indicate the pole
positions including the imaginary parts. The lines show the $\pi\Sigma$,
$\bar{K}N$, $\eta\Lambda$ and $K\Xi$ thresholds.}
 \label{fig:lambdainf}
\end{figure*}

In the fifth energy
level, these two channels become irrelevant, and the coupling strengths to
$\eta\Lambda$ and $K\Xi$ dominate. At the physical point, this level is very
close to the real part of the pole position of the $\Lambda(1670)$ (cf. Table
\ref{tab:coupl}). In the infinite volume, that resonance appears as a quasibound $K\Xi$ state with
relatively small $\bar KN$ and $\pi\Sigma$ branching ratios as Table~\ref{tab:coupl} shows. The overlap with the $\eta \Lambda$ channel is not small
although the branching ratio to this channel is only moderate due to reduced
phase space. In the finite volume, the situation is different because the weight of the $K\Xi$ channel in the wave function, $\tilde P_l$, is reduced as Table~\ref{tab:ps} shows. At higher pion masses, the fifth eigenstate stays close to the non-interacting $\eta\Lambda$ threshold (Fig.~\ref{fig:compajo}), while the pole of the $\Lambda(1670)$ moves considerably away from the $\eta\Lambda$ threshold (Fig.~\ref{fig:lambdainf}). It is thus, not possible to associate the fifth finite-volume eigenstate with the infinite-volume $\Lambda(1670)$ resonance.

We can compare these results with the calculation in the infinite volume using
the formalism described in Section II together with the SU(3) chiral
extrapolation explained before. The  results are shown in Fig.
\ref{fig:lambdainf}. Here, the pole positions in the infinite volume  as a
function of the pion mass are depicted. The four lines represent the
$\pi\Sigma$, $\bar{K}N$, $\eta\Lambda$ and $K\Xi$ thresholds. For masses close
to the physical point, the lowest state is a resonance above the $\pi\Sigma$
threshold. When the mass of the pion increases, the lower state becomes a cusp, i.e., the pole is close to threshold, but on a sheet that is not
directly accessible from the physical axis. When the pion mass increases
further, it becomes a bound state. The second pole of the
$\Lambda(1405)$ is always below and close to the $\bar{K}N$ threshold for all
pion masses considered. A third state appears at higher energies. This state
couples more to the channels $\eta\Lambda$ and $K\Xi$, with larger coupling
strength to $K\Xi$. This state is identified with the $\Lambda(1670)$
\cite{bennhold,Doring:2011ip}.  In Table \ref{tab:couplfi}, we show the 
comparison between the two lowest pole positions and coupling constants in the
infinite and finite volume. For the physical pion mass, we have taken here
larger boxes, $L=4$ fm. In this table, $b_{\bar{K}N}$ and $b_{\pi\Sigma}$ denote
the distances to the $\bar{K}N$ and $\pi\Sigma$ thresholds, where the negative
sign means that the state is above threshold. For masses below $400$ MeV, we
observe that the second state in the finite volume has a dominant
$\bar{K}N$ component, and is between the $\pi\Sigma$ and $\bar{K}N$
thresholds. It shares these properties with the higher-lying pole of the $\Lambda(1405)$ in the infinite volume.  We can understand the proximity of
finite- and infinite-volume states as follows: the second $\Lambda(1405)$ pole
is a quasi- bound $\bar KN$ state with little influence from the $\pi\Sigma$
channel. In the finite volume, the eigenstate appears therefore almost as a
bound state. In the limit of zero $\pi\Sigma$ coupling, the position of finite-
and infinite-volume poles would differ only by exponentially suppressed
corrections scaling with the binding momentum. On the contrary, the lower state
shows very different properties in the infinite volume limit and in the box. As
discussed, for low pion masses, the finite-volume state below the $\pi\Sigma$
threshold is related to the lower $\Lambda(1405)$ pole only insofar, that it
indicates the attractive $\pi\Sigma$ interaction leading to the generation of
the pole in the infinite volume (at a very different position). For high pion
masses, the lower $\Lambda(1405)$ pole in the infinite volume limit becomes a
bound state, and then the couplings  to all the channels become very similar to
the ones in the box as one can see from Table \ref{tab:couplfi}. However, in
this case the masses of the poles are very far away from the lattice
data of Ref. \cite{hall} which can be due to different reasons as discussed before.
\begin{table*}
\begin{center}
{\setlength{\tabcolsep}{0.3em}
 {\renewcommand{\arraystretch}{1.1}
  \begin{tabular}{llrrrrrrrrrr}\hline\hline
  Set&$L(fm)$&$m_\pi$&$m_K$&$m_\eta$&$M_N$&$M_\Lambda$&$M_\Sigma$&$M_\Xi$&$f_\pi$&$f_K$&$f_\eta$\\
  \hline
  1&2.99&170.29&495.78&563.97&962.2&1135.8&1181.5&1323.6&94.5&113.2&122.1\\
  2&3.04&282.84&523.26&581.72&1058.7&1173.4&1235.5&1332.8&102.5&116.1&122.3\\
  3&3.08&387.81&559.46&605.97&1150.1&1261.0&1292.4&1377.4&109.5&118.5&122.6\\
  4&3.23&515.56&609.75&638.07&1274.5&1333.4&1353.5&1401.8&116.3&120.6&122.4\\
  5&3.27&623.14&670.08&685.01&1420.3&1434.2&1449.8&1472.4&120.1&121.9&122.6\\  
  \hline\hline\end{tabular}
}}
 \end{center}
 \caption{Pseudoscalar meson decay constants obtained from SU(3) chiral extrapolation with the masses from \cite{hall}. Units are MeV.}
 \label{tab:par}
\end{table*}

\begin{sidewaystable*}
\vspace{7cm}
\begin{minipage}[rtb]{97mm}
\begin{center}
 {\setlength{\tabcolsep}{0.3em}
 {\renewcommand{\arraystretch}{0.8}
  \begin{tabular}{ccccccc}
  \hline\hline
&$\bar{K} N$&$\pi\Sigma$&$\eta\Lambda$&$K\Xi$&$1-\tilde{Z}$&$\tilde{Z}$\\
\hline
\multicolumn{7}{c}{$W_0=1302(1)$}\\
$\tilde{g}_i$&$-0.69(5)$&$ 0.80(3)  $&$ -0.15(2) $&$0.080(2)$&&\\
$\tilde{P}_l$&$0.020(3)$&$ 0.77(1)$&$   0.0004(1)$&$    0.000090(5)$&$0.79(1)$&$0.21(1)$\\
\multicolumn{7}{c}{$W_0=1394(4)$}\\
$\tilde{g}_i$&$1.9(1)$&$-0.388(5) $&$ 0.94(5) $&$-0.061(6)$&&\\
$\tilde{P}_l$&$0.76(2)$&$ 0.046(6)$&$   0.021(2)$&$    0.00007(1)$&$0.83(2)$&$0.17(2)$\\
\multicolumn{7}{c}{$W_0=1507(4)$}\\
$\tilde{g}_i$&$2.34(2)$&$-0.92(7) $&$ 1.10(3) $&$-0.39(2)$&&\\
$\tilde{P}_l$&$0.68(2)$&$ 0.10(1)$&$   0.046(3)$&$    0.0036(5)$&$0.83(1)$&$0.17(1)$\\
\multicolumn{7}{c}{$W_0=1582(3)$}\\
$\tilde{g}_i$&$0.64(5)$&$1.93(3)$&$0.43(4)$&$0.45(4)$&&\\
$\tilde{P}_l$&$0.09(1)$&$ 0.73(1)$&$  0.014(2)$&$    0.006(1)$&$0.840(3)$&$0.160(3)$\\
\multicolumn{7}{c}{$W_0=1659(2)$}\\
$\tilde{g}_i$&$0.09(1)$&$-0.04(1)$&$-0.3(1)$&$1.3(1)$&&\\
$\tilde{P}_l$&$0.069(6)$&$ 0.002(1)$&$  0.76(5)$&$    0.077(17)$&$0.91(4)$&$0.09(4)$\\
\hline
\multicolumn{7}{c}{$W_0=1322(2)$}\\
$\tilde{g}_i$&$-0.65(5)$&$ 0.90(4)  $&$ -0.14(2) $&$0.105(3)$&&\\
$\tilde{P}_l$&$0.017(3)$&$ 0.80(1)$&$   0.0003(1)$&$    0.00017(1)$&$0.82(1)$&$0.18(1)$\\
\multicolumn{7}{c}{$W_0=1416(4)$}\\
$\tilde{g}_i$&$1.94(13)$&$-0.43(1) $&$ 0.96(5) $&$-0.07(2)$&&\\
$\tilde{P}_l$&$0.76(3)$&$ 0.047(2)$&$   0.021(2)$&$    0.0001(5)$&$0.83(3)$&$0.17(3)$\\
\multicolumn{7}{c}{$W_0=1526(5)$}\\
$\tilde{g}_i$&$2.20(5)$&$-1.14(15)$&$ 1.02(4)$&$-0.5(1)$&&\\
$\tilde{P}_l$&$0.64(5)$&$ 0.16(5)$&$   0.037(3)$&$    0.004(1)$&$0.841(6)$&$0.159(6)$\\
\multicolumn{7}{c}{$W_0=1587(4)$}\\
$\tilde{g}_i$&$0.94(11)$&$1.83(7)$&$0.59(7)$&$0.41(6)$&&\\
$\tilde{P}_l$&$0.15(3)$&$ 0.66(4)$&$  0.019(5)$&$    0.005(1)$&$0.834(7)$&$0.166(7)$\\
\multicolumn{7}{c}{$W_0=1686(5)$}\\
$\tilde{g}_i$&$0.07(4)$&$-0.034(18)$&$-0.59(16)$&$1.8(2)$&&\\
$\tilde{P}_l$&$0.079(8)$&$ 0.009(5)$&$  0.56(11)$&$    0.16(4)$&$0.81(6)$&$0.19(6)$\\
\hline\hline
  \end{tabular}
}}
 \end{center}
\end{minipage}
\begin{minipage}[rtb]{97mm}
 \begin{center}
 {\setlength{\tabcolsep}{0.3em}
 {\renewcommand{\arraystretch}{0.8}
  \begin{tabular}{ccccccc}
  \hline\hline
&$\bar{K} N$&$\pi\Sigma$&$\eta\Lambda$&$K\Xi$&$1-\tilde{Z}$&$\tilde{Z}$\\
\hline
\multicolumn{7}{c}{$W_0=1490(3)$}\\
$\tilde{g}_i$&$-0.85(19)$&$ 1.14(9)  $&$ -0.3(1) $&$0.29(3)$&&\\
$\tilde{P}_l$&$0.05(2)$&$ 0.83(4)$&$   0.003(2)$&$    0.0017(3)$&$0.88(2)$&$0.12(2)$\\
\multicolumn{7}{c}{$W_0=1544(7)$}\\
$\tilde{g}_i$&$1.78(18)$&$-0.31(4)$&$ 1.0(1) $&$-0.14(6)$&&\\
$\tilde{P}_l$&$0.74(8)$&$ 0.09(4)$&$   0.03(5)$&$    0.0006(4)$&$0.86(3)$&$0.14(3)$\\
\multicolumn{7}{c}{$W_0=1633(8)$}\\
$\tilde{g}_i$&$1.85(15)$&$-1.0(2) $&$ 1.13(11)$&$-0.94(3)$&&\\
$\tilde{P}_l$&$0.60(1)$&$ 0.14(6)$&$   0.06(1)$&$    0.03(2)$&$0.83(4)$&$0.17(4)$\\
\multicolumn{7}{c}{$W_0=1693(6)$}\\
$\tilde{g}_i$&$1.00(18)$&$1.72(12)$&$0.43(14)$&$0.66(21)$&&\\
$\tilde{P}_l$&$0.18(6)$&$ 0.63(8)$&$  0.02(1)$&$    0.02(1)$&$0.85(2)$&$0.15(2)$\\
\multicolumn{7}{c}{$W_0=1739(9)$}\\
$\tilde{g}_i$&$0.51(12)$&$-0.3(2)$&$-0.6(2)$&$1.9(3)$&&\\
$\tilde{P}_l$&$0.10(2)$&$ 0.05(5)$&$  0.43(18)$&$    0.20(5)$&$0.78(8)$&$0.22(8)$\\
\hline
\multicolumn{7}{c}{$W_0=1646(8)$}\\
$\tilde{g}_i$&$-1.47(5)$&$ 1.26(12) $&$ -0.8(3) $&$0.51(15)$&&\\
$\tilde{P}_l$&$0.23(12)$&$ 0.59(20)$&$   0.02(2)$&$    0.007(5)$&$0.85(7)$&$0.15(7)$\\
\multicolumn{7}{c}{$W_0=1674(5)$}\\
$\tilde{g}_i$&$1.24(23)$&$0.28(23)$&$ 0.76(15) $&$-1.1(1)$&&\\
$\tilde{P}_l$&$0.56(22)$&$ 0.34(22)$&$   0.021(7)$&$    0.0004(5)$&$0.92(4)$&$0.08(4)$\\
\multicolumn{7}{c}{$W_0=1759(11)$}\\
$\tilde{g}_i$&$1.74(27)$&$-0.86(27)$&$ 1.17(34) $&$-1.1(7)$&&\\
$\tilde{P}_l$&$0.56(11)$&$ 0.12(5)$&$   0.08(3)$&$    0.06(4)$&$0.82(5)$&$0.18(5)$\\
\multicolumn{7}{c}{$W_0=1815(11)$}\\
$\tilde{g}_i$&$1.2(3)$&$1.0(4)$&$-0.29(16)$&$1.9(3)$&&\\
$\tilde{P}_l$&$0.24(1)$&$ 0.33(16)$&$  0.01(1)$&$    0.18(5)$&$0.76(9)$&$0.24(9)$\\
\multicolumn{7}{c}{$W_0=1836(8)$}\\
$\tilde{g}_i$&$0.42(13)$&$-0.8(4)$&$-0.65(16)$&$1.8(3)$&&\\
$\tilde{P}_l$&$0.06(4)$&$ 0.4(2)$&$  0.17(11)$&$    0.19(8)$&$0.82(4)$&$0.18(4)$\\
\hline
\hline

  \end{tabular}
}}
 \end{center}
\end{minipage}
\caption{Pole positions, $W_0$, couplings, $\tilde{g}_i$, $\tilde{P}_l$'s,
$1-\tilde{Z}$ and $\tilde{Z}$ in Eq. (\ref{eq:compwfin}), for the five energy
levels predicted from U$\chi$PT in Fig. \ref{fig:compajo}. From left-top to
bottom, it refers to the physical set and set 1, and from right-top to bottom,
sets 2 and 3 in Table \ref{tab:par}. Pole positions are given in units of MeV.
The errors are obtained by varying subtraction and meson decay
constants as described in Sec.~\ref{sec:spectrum}.}
\label{tab:ps}
\end{sidewaystable*}
 
\begin{table*}
\begin{minipage}[b]{85mm}
\begin{center}
Infinite volume
\vspace{0.2cm}
{\setlength{\tabcolsep}{0.4em}
{\renewcommand{\arraystretch}{1.1}
\begin{tabular}{llrrrrrr}\hline\hline
 Set&& \multicolumn{4}{c}{Channel}\\
&Pole&\multicolumn{4}{c}{$|g_i|$}&$b_{\bar{K}N}$&$b_{\pi\Sigma}$\\
&(MeV)&$\bar{K} N$&$\pi\Sigma$&$\eta\Lambda$&$K\Xi$&(MeV)&(MeV)\\\hline
Phy.&1379-i\,71&2.20&3.1&0.8&0.5&56&-48\\
&1412-i\,19&3.1&1.7&1.5&0.3&23&-81\\
\hline
1&1369-i\,64&1.9&2.9&0.6&0.5&89&-17\\
&1443-i\,17&2.6&1.35&1.32&0.3&15&-91\\
\hline
2&& \multicolumn{4}{c}{Cusp at 1518.34}&64&0\\
&1565-i\,19&2.5&1.5&1.4&0.5&17&-47\\
\hline
3&1671&2.0&1.3&1.1&0.6&39&9\\
&1700-i\,22&2.0&1.6&1.3&0.7&10&-20\\
\hline
4&1836&1.9&1.2&1.7&1.8&48&33\\
&1875-i\,28&1.3&1.8&1.6&1.7&9&-6\\
\hline
5&1998&0.9&0.8&1.9&2.9&92&75\\
&2077-i\,0.5&2.1&0.4&0.3&1.1&13&-4\\
\hline\hline
\end{tabular}}}
\end{center}
\end{minipage}
\begin{minipage}[b]{85mm}
\begin{center}
Finite volume\vspace{0.2cm}\\
{\setlength{\tabcolsep}{0.4em}
{\renewcommand{\arraystretch}{1.1}
\begin{tabular}{llrrrrrr}\hline\hline
 & \multicolumn{4}{c}{Channel}\\
Pole&\multicolumn{4}{c}{$|\tilde{g}_i|$}&$b_{\bar{K}N}$&$b_{\pi\Sigma}$\\
(MeV)&$\bar{K} N$&$\pi\Sigma$&$\eta\Lambda$&$K\Xi$&(MeV)&(MeV)\\\hline
1322&0.5&0.6&0.1&0.07&113&9\\
1401&2.2&1.0&1.0&0.2&34&-70\\
\hline
1322&0.6&0.9&0.1&0.1&136&30\\
1417&1.9&0.4&0.9&0.06&41&-65\\
\hline
1489&0.9&1.2&0.3&0.3&93&29\\
1541&1.9&0.3&1.0&0.2&41&-23\\
\hline
1649&1.4&1.3&0.7&0.5&61&31\\
1676&1.5&0.1&0.9&0.06&34&4\\
\hline
1829&1.8&1.2&1.5&1.5&55&40\\
1859&0.9&0.5&0.6&0.09&25&10\\
\hline
1997&1.0&0.9&1.9&2.9&93&76\\
2062&0.9&0.7&0.2&0.9&28&11\\
\hline\hline
\end{tabular}}}
\end{center}
\end{minipage}\vspace{0.2cm}
\caption{Pole positions and couplings of the states $|g_i|$ in the infinite
(left, A) and finite (right, B) volume for all sets. Note that for the physical set in
the finite volume (``Phy.'') we have taken larger volumes ($L=4$ fm) for the
values that are shown in this table. For the other sets of pion masses of Ref.
\cite{hall}, $L\simeq 3$ fm is used. It should be stressed that finite- and
infinite-volume poles cannot be directly identified with each other.}
\label{tab:couplfi}
\end{table*}

\begin{figure*}[h]\begin{center}
 \includegraphics[width=1.0\linewidth]{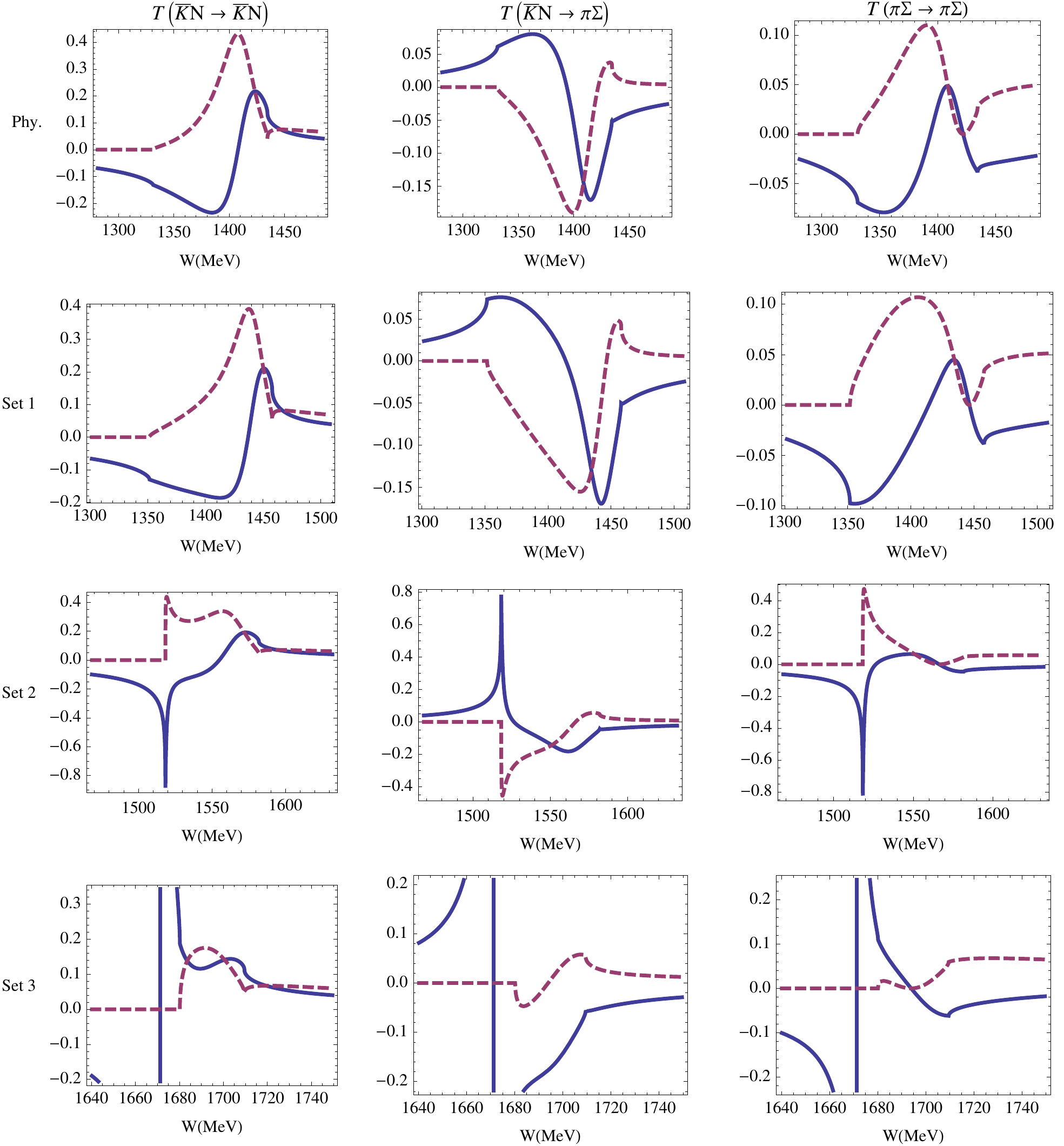}
 \end{center}
 \caption{Real (solid lines) and imaginary parts (dashed lines) of the
meson-baryon scattering amplitude of the four coupled channels $\bar{K}N$,
$\pi\Sigma$, $\eta\Lambda$ and $K\Xi$. From top to bottom, the figures
correspond to physical pion mass and sets 1 to 3 of Table \ref{tab:par}. The
left column corresponds to the $T_{\bar{K}N\to\bar{K}N}$ amplitude, the central
column corresponds to the transition $T_{\bar{K}N\to\pi\Sigma}$ and the right
column to $T_{\pi\Sigma\to\pi\Sigma}$.}
 \label{fig:lambdaam}
\end{figure*} 


\subsection*{Meson-baryon scattering amplitudes for different pion masses in the
infinite volume limit}
Finally, we provide the infinite-volume scattering amplitudes for
strangeness=$-1$ in Fig. \ref{fig:lambdaam}. Every row shows the real (solid
lines) and imaginary part (dashed lines) of the scattering amplitude,
$T_{\bar{K}N\to\bar{K}N},T_{\bar{K}N\to\pi\Sigma}$ and
$T_{\pi\Sigma\to\pi\Sigma}$. The first row shows the amplitude for the physical
pion mass, while the second to fourth rows correspond to pion masses of $170$,
$282$ and $388$ MeV (first three sets in Table \ref{tab:par}). For the physical
set, these amplitudes are very similar to the ones obtained in the work of Ref.
\cite{bennhold}, where we observe the presence of two resonances related to the
two poles near the energy of the $\Lambda(1405)$. For higher pion masses the
lighter pole of the $\Lambda(1405)$ first becomes a cusp (third row) and then a
bound state (fourth row). The heavier pole of the $ \Lambda(1405)$ couples
predominantly to the $\bar K N$ channel as the figure in the upper left corner
shows. As the pion mass increases, the pole remains close to the $\bar K N$
threshold as a quasi-bound state.  

The fact that the second pole of the $\Lambda(1405)$ always appears close
to the $\bar{K}N$ threshold may be due to that the kaon mass, which controls the
strength of the $\bar{K}N\to\bar{K}N$ Weinberg-Tomozawa term, does not change much, so that the
properties of the bound state also do not experience much variations. In
contrast, the $\pi\Sigma\to\pi\Sigma$ Weinberg-Tomozawa
interaction becomes significantly stronger with increasing pion mass, changing drastically the nature of the
lower state from resonance to bound state.


\section{Conclusions}

The quark mass dependence of the energy levels in a box for the coupled channels
with $J^P=\frac{1}{2}^-, I=0, S=-1$ has been studied, using the Weinberg-Tomozawa term from the lowest order
 $\chi$PT interaction. This dependence has been compared to the lattice data
of Ref.~\cite{hall} and extrapolated to the infinite volume. U$\chi$PT predicts a
two-pole structure for the $\Lambda(1405)$. In the finite volume, two energy
levels close to the $\pi\Sigma$ and $\bar{K}N$ thresholds are found.  The second
energy level agrees well with the lattice data of Ref. \cite{hall} for pion
masses below $400$ MeV, in the estimated limit of applicability of the present approach.
This energy level shows a large coupling and overlap with the $\bar{K}N$ channel
and has similar properties as the higher pole of the $\Lambda(1405)$. The state
remains quasi-bound in the $\bar K N$ channel and close to its threshold, as the
pion mass increases. Thus, the lattice data of Ref. \cite{hall} are 
not in contradiction
with the two-pole hypothesis for the $\Lambda(1405)$. Yet, these data proof by
no means that hypothesis. For this, a few remaining obstacles need to be
addressed: The first problem is the absence of the $\pi\Sigma$ threshold level
in the lattice calculation of Ref. \cite{hall}, that appears here below  the
$\pi\Sigma$ threshold, indicating an attractive $\pi\Sigma$ interaction. In the
infinite volume, this attraction leads to the generation of a second (lighter)
pole of the $\Lambda(1405)$. This behavior is universal in U$\chi$PT
calculations, and also present in some dynamical coupled-channel
approaches. Here, we have assumed that this absence is due to the absence of meson-baryon operators in the operator base used in Ref.~ \cite{hall}. To propose
suitable meson-baryon operators for the detection of the threshold level, we
have considered the finite-volume analog of $\tilde{P}_l$ that specify the
relative weight of a channel in a state's wave function. It turns out that an
operator of the $\pi\Sigma$ type is most suited to detect the level in future
lattice simulations. Indeed, the precise location of that level would specify
the size of attraction in the $\pi\Sigma$ channel at threshold and help to pin down
the location of the lighter $\Lambda(1405)$ pole that is notoriously difficult
to determine. Also, a precise determination of the pole positions from lattice data requires to populate the region between the
$\pi\Sigma$ and $\bar K N$ thresholds with more lattice eigenvalues, using,
e.g., moving frames and asymmetric boxes
\cite{MartinezTorres:2012yi}. Also, higher-order U$\chi$PT calculations along
the lines of Refs. \cite{Guo:2012vv, Ikeda:2012au, Mai:2012dt} will be needed to assess
theoretical uncertainties in direct fits to future lattice data.


\acknowledgements
We gratefully acknowledge support from the NSF/PIF award no. 1415459, the
NSF/Career award no. 1452055 and a GWU startup grant. M. D. is also supported by the U.S. Department of Energy, Office of Science, Office of Nuclear Physics under Contract No. DE-AC05-06OR23177. We also thank D. Leinweber, M. Mai, E. Oset, and R. D. Young for clarifying discussions and J.M.M. Hall for
providing details of the lattice calculation.


\section*{Appendix\\\vspace{0.3cm} SU(3) chiral extrapolation}
Chiral symmetry is explicitly broken and gives rise to masses of the quarks
$u,d,s$ different from zero. Then, the Goldstone bosons acquire masses which at
leading order are related to the chiral condensate and are denoted here as
$M_{0\pi}$, $M_{0K}$ and $M_{0\eta}$. To one loop, the masses of the  Goldstone
bosons carry corrections and the physical masses can be expressed as a function
of the leading order masses ($M_0$), LEC's ($L^r$) and pseudoscalar decay
constants ($f$). The following formulas for the pseudoscalar masses, derived
from the SU(3) chiral extrapolation are taking from Ref. \cite{jenifer}, which is
based on chiral perturbation theory \cite{Gasser:1984gg},
\begin{eqnarray}
&&M_\pi^2= M_{0\,\pi}^2\left[1+\mu_\pi-\frac{\mu_\eta}{3}+\frac{16
M_{0\,K}^2}{f_0^2}\left(2L_6^r-L_4^r\right)\right.\nonumber\\&&
+\left.
\frac{8
M_{0\,\pi}^2}{f_0^2}\left(2L_6^r+2L_8^r-L_4^r-L_5^r\right)
\right], \label{pimass}
\end{eqnarray}
 \begin{eqnarray}
&&M^2_K= M^2_{0\,K}\left[1+\frac{2\mu_\eta}{3}+\frac{8
M_{0\,\pi}^2}{f_0^2}\left(2L_6^r-L_4^r\right)\right.\nonumber\\&&
+\left.\frac{8
M_{0\,K}^2}{f_0^2}\left(4L_6^r+2L_8^r-2L_4^r-L_5^r\right)\right],
\label{kmass}
\end{eqnarray}
\begin{eqnarray}
&&M^2_\eta= M^2_{0\,\eta} \left[1+2\mu_K-\frac{4}{3}\mu_\eta+
\frac{8M^2_{0\,\eta}}{f_0^2}(2L_8^r-L_5^r)\right.\nonumber\\&&
+\left.
\frac{8}{f_0^2}(2 M^2_{0\,K}+M^2_{0\,\pi})(2L_6^r-L_4^r)
\right]\nonumber\\
&&+ M^2_{0\,\pi}\left[-\mu_\pi+\frac{2}{3}\mu_K+\frac{1}{3}\mu_\eta\right]
\nonumber\\&&+\frac{128}{9f_0^2}(M^2_{0\,K}-M^2_{0\,\pi})^2(3L_7+L_8^r),
 \label{etamass}
\end{eqnarray}
with
\begin{eqnarray}
&&\hspace{0.5cm}\mu_P=\frac{M_{0\, P}^2}{32 \pi^2 f_0^2}
\log\frac{M_{0 \,P}^2}{\mu^2},
\qquad P=\pi,K,\eta\ .
\end{eqnarray}
In the above equations $f_0$ is the pion decay constant in the chiral limit,
$4\pi f_0\simeq 1.2$ GeV, $\mu$ is the  regularization scale, commonly fixed at
$\mu=M_\rho$, and $L^r_i$'s, with $i=1,8$, are the  Low Energy Constants which
multiply the tree level diagrams of $\mathcal{O}(p^4)$ present in the  next to
leading order $t_4(s)$ term in the $\chi PT$ expansion of the amplitude for
meson-meson scattering ($t(s)=t_2(s)+t_4(s)+...$,
$t_{2k}=\mathcal{O}(p^{2k})$). 

On the other hand, the decay constants evaluated to one loop SU(3) $\chi PT$,
are expressed in terms of the masses at leading order as 
\begin{eqnarray}
&&f_\pi= f_0\left[1-2\mu_\pi-\mu_K+\frac{4
M_{0\,\pi}^2}{f_0^2}\left(L_4^r+L_5^r\right)+\frac{8
M_{0\,K}^2}{f_0^2}L_4^r\right], \nonumber\\\label{fpis}\\ &&f_K=
f_0\left[1-\frac{3\mu_\pi}{4}-\frac{3\mu_K}{2}-\frac{3\mu_\eta}{4}+\frac{4
M_{0\,\pi}^2}{f_0^2}L_4^r\right.\nonumber\\&&+\left.\frac{4
M_{0\,K}^2}{f_0^2}\left(2L_4^r+L_5^r\right)\right],
 \label{fk}\\
&&f_\eta=f_0\left[1-3\mu_K+ \frac{4
L_4^r}{f_0^2}\left(M_{0\,\pi}^2+2M_{0\,K}^2\right)
+\frac{4M_{0\,\eta}^2}{f_0^2}L_5^r\right]\ .\nonumber\\
\label{feta}
\end{eqnarray}

The $L^r_i$'s values used here are taken from Fit I of Ref. \cite{jenifer} to
experiment and lattice data (shown in Table 1 of Ref. \cite{jenifer}).

In order to evaluate the meson decay constants for the different sets of Table
\ref{tab:par}, first  Eqs. (\ref{pimass}), (\ref{kmass}), (\ref{fpis}) and
(\ref{fk}) are evaluated at the physical point (Table \ref{tab:par}), obtaining
the  values of the four variables, $M_{0\pi}$, $M_{0K}$ $M_{0\eta}$ and $f_0$.
In these sets of equations, the LEC $L^r_7$ does not appear. Once these
constants are known, $L^r_7$ is fixed to obtain the mass of the $\eta$ at the
physical point, given by Eq. (\ref{etamass}). This gives as a result,
$L^r_7=-0.423\times 10^{-3}$,  very close to the one obtained in \cite{jenifer}
($-0.43\times 10^{-3}$). For $f_0$, a value of $79.2$ is obtained, and using the
formulas of Eqs. (\ref{pimass})-(\ref{feta}), we evaluate the $M_0$'s and $f$'s
for $\pi$, $K$ and $\eta$ for every set of masses in Table \ref{tab:par}. The
decay constants obtained are shown in Table \ref{tab:par} of the Results
Section.


\begin{thebibliography}{99}

\bibitem{Isgur:1978xj} 
  N.~Isgur and G.~Karl,
  Phys.\ Rev.\ D {\bf 18}, 4187 (1978).
\bibitem{Inoue:2006nf} 
  T.~Inoue,
  Nucl.\ Phys.\ A {\bf 790}, 530 (2007).
\bibitem{Dalitz:1960du} 
  R.~H.~Dalitz and S.~F.~Tuan,
  Annals Phys.\  {\bf 10}, 307 (1960).
\bibitem{Dalitz:1967fp} 
  R.~H.~Dalitz, T.~C.~Wong and G.~Rajasekaran,
  Phys.\ Rev.\  {\bf 153}, 1617 (1967).
\bibitem{Veit:1984jr} 
  E.~A.~Veit, B.~K.~Jennings, A.~W.~Thomas and R.~C.~Barrett,
  Phys.\ Rev.\ D {\bf 31}, 1033 (1985).

\bibitem{Siegel:1994mb} 
  P.~B.~Siegel and B.~Saghai,
  Phys.\ Rev.\ C {\bf 52}, 392 (1995).
\bibitem{Tanaka:1992gj} 
  K.~Tanaka and A.~Suzuki,
  Phys.\ Rev.\ C {\bf 45}, 2068 (1992).
\bibitem{Chao:1973sa} 
  Y.~A.~Chao, R.~W.~Kraemer, D.~W.~Thomas and B.~R.~Martin,
  Nucl.\ Phys.\ B {\bf 56}, 46 (1973).
\bibitem{confedalitz}
R. H. Dalitz and J. G. McGinley, Proceedings of the international Conference on Hypernuclear and Kaon Physics (North Holland, Heidelberg, 1982)

\bibitem{Kamano:2014zba} 
  H.~Kamano, S.~X.~Nakamura, T.-S.~H.~Lee and T.~Sato,
  Phys.\ Rev.\ C {\bf 90}, 065204 (2014).

\bibitem{Kamano:2015hxa} 
  H.~Kamano, S.~X.~Nakamura, T.-S.~H.~Lee and T.~Sato,
  Phys.\ Rev.\ C {\bf 92}, no. 2, 025205 (2015).


\bibitem{Zhang:2013cua} 
  H.~Zhang, J.~Tulpan, M.~Shrestha and D.~M.~Manley,
  Phys.\ Rev.\ C {\bf 88}, 035204 (2013).
  
\bibitem{Fernandez-Ramirez:2015tfa} 
  C.~Fern\'andez-Ram\'irez, I.~V.~Danilkin, D.~M.~Manley, V.~Mathieu and A.~P.~Szczepaniak,
  Phys.\ Rev.\ D {\bf 93}, no. 3, 034029 (2016).

\bibitem{Jackson:2015dva} 
  B.~C.~Jackson, Y.~Oh, H.~Haberzettl and K.~Nakayama,
  Phys.\ Rev.\ C {\bf 91}, 065208 (2015).

\bibitem{magasan} 
  A.~Feijoo, V.~K.~Magas and A.~Ramos,
  Phys.\ Rev.\ C {\bf 92}, 015206 (2015).

\bibitem{Briscoe:2015qia} 
 W.~J.~Briscoe, M.~D\"oring, H.~Haberzettl, D.~M.~Manley, M.~Naruki, I.~I.~Strakovsky and E.~S.~Swanson,
  Eur.\ Phys.\ J.\ A {\bf 51}, 129 (2015).

\bibitem{Gasser:2007zt} 
 J.~Gasser, V.~E.~Lyubovitskij and A.~Rusetsky,
  Phys.\ Rept.\  {\bf 456}, 167 (2008).

\bibitem{Jennings:1986yg} 
  B.~K.~Jennings,
  Phys.\ Lett.\ B {\bf 176}, 229 (1986).
  
\bibitem{Arima:1990yv} 
  M.~Arima and K.~Yazaki,
  Nucl.\ Phys.\ A {\bf 506}, 553 (1990).
  
\bibitem{Arima:1994bv} 
  M.~Arima, S.~Matsui and K.~Shimizu,
  Phys.\ Rev.\ C {\bf 49}, 2831 (1994).
  
\bibitem{He:1993et} 
  G.~l.~He and R.~H.~Landau,
  Phys.\ Rev.\ C {\bf 48}, 3047 (1993).
  
\bibitem{Kumar:1980hs} 
  K.~S.~Kumar and Y.~Nogami,
  Phys.\ Rev.\ D {\bf 21}, 1834 (1980).
  
\bibitem{Schnick:1987is} 
  J.~Schnick and R.~H.~Landau,
  Phys.\ Rev.\ Lett.\  {\bf 58}, 1719 (1987).
  
\bibitem{Fink:1989uk} 
  P.~J.~Fink, Jr., G.~He, R.~H.~Landau and J.~W.~Schnick,
  Phys.\ Rev.\ C {\bf 41}, 2720 (1990).

\bibitem{Iwasaki:1997wf} 
  M.~Iwasaki, R.~S.~Hayano, T.~M.~Ito, S.~N.~Nakamura, T.~P.~Terada, D.~R.~Gill, L.~Lee and A.~Olin {\it et al.},
  Phys.\ Rev.\ Lett.\  {\bf 78}, 3067 (1997).
  
\bibitem{Ito:1998yi} 
  T.~M.~Ito, R.~S.~Hayano, S.~N.~Nakamura, T.~P.~Terada, M.~Iwasaki, D.~R.~Gill, L.~Lee and A.~Olin {\it et al.},
  Phys.\ Rev.\ C {\bf 58}, 2366 (1998).





\bibitem{Deser:1954vq} 
  S.~Deser, M.~L.~Goldberger, K.~Baumann and W.~E.~Thirring,
  Phys.\ Rev.\  {\bf 96}, 774 (1954).


\bibitem{Meissner:2004jr} 
  U.-G.~Mei{\ss}ner, U.~Raha and A.~Rusetsky,
  Eur.\ Phys.\ J.\ C {\bf 35}, 349 (2004).


\bibitem{Bazzi:2011zj} 
  M.~Bazzi, G.~Beer, L.~Bombelli, A.~M.~Bragadireanu, M.~Cargnelli, G.~Corradi, C.~Curceanu (Petrascu) and A.~d'Uffizi {\it et al.},
  Phys.\ Lett.\ B {\bf 704}, 113 (2011).


\bibitem{DEARSID} 
  G.~Beer {\it et al.} [DEAR Collaboration],
  Phys.\ Rev.\ Lett.\  {\bf 94}, 212302 (2005).

\bibitem{Ikeda:2012au} 
  Y.~Ikeda, T.~Hyodo and W.~Weise,
  Nucl.\ Phys.\ A {\bf 881}, 98 (2012).

\bibitem{Borasoy:2005ie} 
 B.~Borasoy, R.~Ni{\ss}ler and W.~Weise,
  Eur.\ Phys.\ J.\ A {\bf 25}, 79 (2005).

\bibitem{Oller:2005ig} 
  J.~A.~Oller, J.~Prades and M.~Verbeni,
  Phys.\ Rev.\ Lett.\  {\bf 95}, 172502 (2005).
  
\bibitem{Oller:2006jw} 
  J.~A.~Oller,
  Eur.\ Phys.\ J.\ A {\bf 28}, 63 (2006)

 \bibitem{Mai:2012dt} 
  M.~Mai and U.~G.~Mei{\ss}ner,
  Nucl.\ Phys.\ A {\bf 900}, 51  (2013).

\bibitem{Guo:2012vv} 
  Z.~H.~Guo and J.~A.~Oller,
  Phys.\ Rev.\ C {\bf 87}, 035202 (2013).

  
\bibitem{Kamiya:2016jqc} 
Y.~Kamiya, K.~Miyahara, S.~Ohnishi, Y.~Ikeda, T.~Hyodo, E.~Oset and W.~Weise,
  arXiv:1602.08852 [hep-ph].

\bibitem{Cieply:2016jby} 
  A.~Ciepl\'y, M.~Mai, U.-G.~Mei{\ss}ner and J.~Smejkal,
  arXiv:1603.02531 [hep-ph].
 
\bibitem{Oller:2000fj} 
  J.~A.~Oller and U.-G.~Mei{\ss}ner,
  Phys.\ Lett.\ B {\bf 500}, 263 (2001).
  
\bibitem{Doring:2011xc} 
  M.~D\"oring and U.-G.~Mei{\ss}ner,
  Phys.\ Lett.\ B {\bf 704}, 663 (2011).
  
  
\bibitem{Mai:2014uma} 
  M.~Mai, V.~Baru, E.~Epelbaum and A.~Rusetsky,
  Phys.\ Rev.\ D {\bf 91}, 054016 (2015).

\bibitem{Kaiser:1996js} 
  N.~Kaiser, T.~Waas and W.~Weise,
  Nucl.\ Phys.\ A {\bf 612}, 297 (1997).

  \bibitem{bennhold}
  E.~Oset, A.~Ramos and C.~Bennhold,
  Phys.\ Lett.\ B {\bf 527}, 99 (2002).

\bibitem{Kaiser:1995eg}
N. Kaiser, P. B. Siegel and W. Weise,
Nucl. Phys. {\bf A594}, 325 (1995).

\bibitem{Oset:1997it} 
  E.~Oset and A.~Ramos,
  Nucl.\ Phys.\ A {\bf 635}, 99 (1998).

\bibitem{Jido:2002yz} 
  D.~Jido, A.~Hosaka, J.~C.~Nacher, E.~Oset and A.~Ramos,
  Phys.\ Rev.\ C {\bf 66}, 025203 (2002).

\bibitem{Jido:2003cb}
  D.~Jido, J.~A.~Oller, E.~Oset, A.~Ramos and U.-G.~Mei\ss ner,
  Nucl.\ Phys.\ A {\bf 725}, 181 (2003).

   
  \bibitem{GarciaRecio:2002td} 
  C.~Garcia-Recio, J.~Nieves, E.~Ruiz Arriola and M.~J.~Vicente Vacas,
  Phys.\ Rev.\ D {\bf 67}, 076009 (2003).

\bibitem{Doring:2010rd} 
  M.~D\"oring, D.~Jido and E.~Oset,
  Eur.\ Phys.\ J.\ A {\bf 45}, 319 (2010).

\bibitem{Borasoy:2006sr} 
 B.~Borasoy, U.-G.~Mei{\ss}ner and R.~Ni{\ss}ler,
  Phys.\ Rev.\ C {\bf 74}, 055201 (2006).

  \bibitem{Geng:2007vm} 
  L.~S.~Geng and E.~Oset,
  Eur.\ Phys.\ J.\ A {\bf 34}, 405 (2007).

\bibitem{Magas:2005vu}
  V.~K.~Magas, E.~Oset and A.~Ramos,
  Phys.\ Rev.\ Lett.\  {\bf 95}, 052301 (2005).
  
  \bibitem{Prakhov:2004an} 
  S.~Prakhov {\it et al.}  [Crystall Ball Collaboration],
  Phys.\ Rev.\ C {\bf 70}, 034605 (2004).

\bibitem{Geng:2007hz} 
  L.~S.~Geng, E.~Oset and M.~D\"oring,
  Eur.\ Phys.\ J.\ A {\bf 32}, 201 (2007).

\bibitem{Hyodo:2003jw} 
  T.~Hyodo, A.~Hosaka, E.~Oset, A.~Ramos and M.~J.~Vicente Vacas,
  Phys.\ Rev.\ C {\bf 68}, 065203 (2003).
  
\bibitem{Hyodo:2004vt}
  T.~Hyodo, A.~Hosaka, M.~J.~Vicente Vacas and E.~Oset,
  Phys.\ Lett.\  B {\bf 593}, 75 (2004).

  \bibitem{Jido:2009jf}
  D.~Jido, E.~Oset and T.~Sekihara,
  Eur.\ Phys.\ J.\  A {\bf 42}, 257 (2009).

\bibitem{Sekihara:2014kya} 
  T.~Sekihara, T.~Hyodo and D.~Jido,
  PTEP {\bf 2015}, 063D04 (2015).

\bibitem{Hyodo:2008xr}
  T.~Hyodo, D.~Jido and A.~Hosaka,
  Phys.\ Rev.\  C {\bf 78}, 025203 (2008).

\bibitem{Fernandez-Ramirez:2015fbq} 
 C.~Fern\'andez-Ram\'irez, I.~V.~Danilkin, V.~Mathieu and A.~P.~Szczepaniak,
  Phys.\ Rev.\ D {\bf 93}, 074015 (2016).

\bibitem{Moriya:2013eb} 
  K.~Moriya {\it et al.}  [CLAS Collaboration],
  Phys.\ Rev.\ C {\bf 87}, 035206 (2013).

\bibitem{Moriya:2014kpv} 
  K.~Moriya {\it et al.}  [CLAS Collaboration],
  Phys.\ Rev.\ Lett.\  {\bf 112}, 082004 (2014).

\bibitem{Mai:2014xna} 
 M.~Mai and U.-G.~Mei{\ss}ner,
  Eur.\ Phys.\ J.\ A {\bf 51}, 30 (2015).

\bibitem{Roca:2013av} 
  L.~Roca and E.~Oset,
  Phys.\ Rev.\ C {\bf 87}, 055201 (2013)

\bibitem{Roca:2013cca} 
  L.~Roca and E.~Oset,
  Phys.\ Rev.\ C {\bf 88}, 055206 (2013).

\bibitem{Doring:2011ip} 
  M.~D\"oring, J.~Haidenbauer, U.-G.~Mei{\ss}ner and A.~Rusetsky,
  Eur.\ Phys.\ J.\ A {\bf 47}, 163 (2011).

\bibitem{Lage:2009zv} 
  M.~Lage, U.-G.~Mei{\ss}ner and A.~Rusetsky,
  Phys.\ Lett.\ B {\bf 681}, 439 (2009).
 
\bibitem{Doring:2013glu} 
  M.~D\"oring, M.~Mai and U.-G.~Mei{\ss}ner,
  Phys.\ Lett.\ B {\bf 722}, 185 (2013).

\bibitem{Liu:2016wxq} 
  Z.~W.~Liu, J.~M.~M.~Hall, D.~B.~Leinweber, A.~W.~Thomas and J.~J.~Wu,
  arXiv:1607.05856 [nucl-th].


\bibitem{Bulava:2010yg} 
  J.~Bulava, R.~G.~Edwards, E.~Engelson, B.~Joo, H.~W.~Lin, C.~Morningstar, D.~G.~Richards and S.~J.~Wallace,
  Phys.\ Rev.\ D {\bf 82}, 014507 (2010).

\bibitem{Menadue:2011pd} 
 B.~J.~Menadue, W.~Kamleh, D.~B.~Leinweber and M.~S.~Mahbub,
  Phys.\ Rev.\ Lett.\  {\bf 108}, 112001 (2012).

\bibitem{Engel:2012qp} 
  G.~P.~Engel {\it et al.}  [BGR (Bern-Graz-Regensburg) Collaboration],
  Phys.\ Rev.\ D {\bf 87}, 034502 (2013).

\bibitem{Engel:2013ig} 
  G.~P.~Engel {\it et al.}  [BGR Collaboration],
  Phys.\ Rev.\ D {\bf 87}, 074504 (2013).

\bibitem{Edwards:2012fx} 
  R.~G.~Edwards {\it et al.}  [Hadron Spectrum Collaboration],
  Phys.\ Rev.\ D {\bf 87}, 054506 (2013).

\bibitem{Melnitchouk:2002eg} 
  W.~Melnitchouk, S.~O.~Bilson-Thompson, F.~D.~R.~Bonnet, J.~N.~Hedditch, F.~X.~Lee, D.~B.~Leinweber, A.~G.~Williams and J.~M.~Zanotti {\it et al.},
  Phys.\ Rev.\ D {\bf 67}, 114506 (2003).
  
\bibitem{WalkerLoud:2008bp} 
  A.~Walker-Loud, H.-W.~Lin, D.~G.~Richards, R.~G.~Edwards, M.~Engelhardt, G.~T.~Fleming, P.~Hagler and B.~Musch {\it et al.},
  Phys.\ Rev.\ D {\bf 79}, 054502 (2009).

\bibitem{Lang:2012db} 
  C.~B.~Lang and V.~Verduci,
  Phys.\ Rev.\ D {\bf 87}, no. 5, 054502 (2013).

  \bibitem{hall} 
  J.~M.~M.~Hall, W.~Kamleh, D.~B.~Leinweber, B.~J.~Menadue, B.~J.~Owen, A.~W.~Thomas and R.~D.~Young,
  Phys.\ Rev.\ Lett.\  {\bf 114}, 132002 (2015).

\bibitem{Tovee:1971ga}
D. N. Tovee {\it et al.},
Nucl. Phys. {\bf B33}, 493 (1971).

\bibitem{Nowak:1978au}
R. J. Nowak {\it et al.},
Nucl. Phys. {\bf B139},61 (1978).

\bibitem{jenifer} 
  J.~Nebreda and J.~R.~Pelaez.,
  Phys.\ Rev.\ D {\bf 81}, 054035 (2010).

  
\bibitem{Weinberg:1962hj} 
  S.~Weinberg,
  Phys.\ Rev.\  {\bf 130}, 776 (1963).

\bibitem{Weinberg:1965zz} 
  S.~Weinberg,
  Phys.\ Rev.\  {\bf 137}, B672 (1965).

\bibitem{Baru:2003qq} 
  V.~Baru, J.~Haidenbauer, C.~Hanhart, Y.~Kalashnikova and A.~E.~Kudryavtsev,
  Phys.\ Lett.\ B {\bf 586}, 53 (2004)
  \bibitem{Hanhart:2011jz} 
  C.~Hanhart, Y.~S.~Kalashnikova and A.~V.~Nefediev,
  Eur.\ Phys.\ J.\ A {\bf 47}, 101 (2011)

\bibitem{Sekihara:2010uz} 
  T.~Sekihara, T.~Hyodo and D.~Jido,
  Phys.\ Rev.\ C {\bf 83}, 055202 (2011). 

\bibitem{Hyodo:2011qc} 
  T.~Hyodo, D.~Jido and A.~Hosaka,
  Phys.\ Rev.\ C {\bf 85}, 015201 (2012).

\bibitem{Hyodo:2013nka} 
  T.~Hyodo,
  Int.\ J.\ Mod.\ Phys.\ A {\bf 28}, 1330045 (2013).

\bibitem{danioset} 
  D.~Gamermann, J.~Nieves, E.~Oset and E.~Ruiz Arriola,
  Phys.\ Rev.\ D {\bf 81}, 014029 (2010)

\bibitem{acetioset} 
  F.~Aceti, L.~R.~Dai, L.~S.~Geng, E.~Oset and Y.~Zhang,
  Eur.\ Phys.\ J.\ A {\bf 50}, 57 (2014).

\bibitem{Aceti:2012dd} 
  F.~Aceti and E.~Oset,
  Phys.\ Rev.\ D {\bf 86}, 014012 (2012)

\bibitem{ollerguo} 
  Z.~H.~Guo and J.~A.~Oller,
  Phys.\ Rev.\ D {\bf 93}, 096001 (2016).

\bibitem{Garcia-Recio:2015jsa} 
  C.~Garcia-Recio, C.~Hidalgo-Duque, J.~Nieves, L.~L.~Salcedo and L.~Tolos,
  Phys.\ Rev.\ D {\bf 92}, 034011 (2015).

\bibitem{Sekihara:2015gvw} 
 T.~Sekihara, T.~Arai, J.~Yamagata-Sekihara and S.~Yasui,
   Phys.\ Rev.\ C {\bf 93}, 035204 (2016).

\bibitem{Oller:1998hw} 
  J.~A.~Oller, E.~Oset and J.~R.~Pelaez,
  Phys.\ Rev.\ D {\bf 59}, 074001 (1999)
  [Erratum-ibid.\ D {\bf 60}, 099906 (1999)]
  [Erratum-ibid.\ D {\bf 75}, 099903 (2007)]
  [hep-ph/9804209].

  \bibitem{Doring:2011vk} 
  M.~D\"oring, U.-G.~Mei{\ss}ner, E.~Oset and A.~Rusetsky,
  Eur.\ Phys.\ J.\ A {\bf 47}, 139 (2011).

\bibitem{MartinezTorres:2011pr} 
  A.~Mart\'inez Torres, L.~R.~Dai, C.~Koren, D.~Jido and E.~Oset,
  Phys.\ Rev.\ D {\bf 85}, 014027 (2012).

\bibitem{MartinezTorres:2012yi} 
A.~Martinez Torres, M.~Bayar, D.~Jido and E.~Oset,
  Phys.\ Rev.\ C {\bf 86}, 055201 (2012).

\bibitem{Gasser:1984gg}
  J.~Gasser and H.~Leutwyler,
  Nucl.\ Phys.\  B {\bf 250}, 465 (1985).

\end{thebibliography}
\end{document}